\documentclass[12pt]{iopart}

\usepackage{graphicx}
\usepackage{bm} 
\usepackage{subfigure}
\usepackage[colorlinks=true,linktocpage=true,linkcolor=blue,citecolor=blue]{hyperref}
\usepackage[usenames,dvipsnames]{color}

\begin{document}

\title[Exact solutions of the (0+1)-dimensional Boltzmann equation]{
Exact solution of the (0+1)-dimensional \\ Boltzmann equation for massive \\ Bose-Einstein and Fermi-Dirac gases}

\author{Wojciech Florkowski}
\address{ $^1$Institute of Physics, Jan Kochanowski University, PL-25406~Kielce, Poland}
\address{ $^2$The H. Niewodnicza\'nski Institute of Nuclear Physics, Polish Academy of Sciences, PL-31342 Krak\'ow, Poland} 
\ead{Wojciech.Florkowski@ifj.edu.pl}

\author{Ewa Maksymiuk}
\address{Institute of Physics, Jan Kochanowski University, PL-25406~Kielce, Poland}
\ead{MaksymiukEwa@gmail.com}

\begin{abstract}
We present the exact solution of the (0+1)-dimensional Boltzmann equation for massive Bose-Einstein and Fermi-Dirac gases. For the initial conditions used typically in ultra-relativistic heavy-ion collisions, we find that the effects of quantum statistics are very small for thermodynamics-like functions such as the effective temperature, energy density or transverse and longitudinal pressures.  Similarly, the inclusion of quantum statistics affects very little the shear viscosity. On the other hand, the quantum statistics becomes important for description of the phenomena connected with the bulk viscosity.
\end{abstract}

%Uncomment for PACS numbers title message
% Keywords required only for MST, PB, PMB, PM, JOA, JOB? 
\vspace{2pc}

% Uncomment for Submitted to journal title message
\submitto{\JPG}
% Comment out if separate title page not required
\maketitle

%%%%%%%%%%%%%%%%%%%%%%%%%%%%%%%%%%%%%%%%%%%%%%%%%%%%%%%%%%%%%%%%%%%%%%%%%%%%%%%%%%%%%%%%%%%%%%%%%%%%
\section{Introduction}
\label{sect:intro}
%%%%%%%%%%%%%%%%%%%%%%%%%%%%%%%%%%%%%%%%%%%%%%%%%%%%%%%%%%%%%%%%%%%%%%%%%%%%%%%%%%%%%%%%%%%%%%%%%%

In this paper we generalize recent results describing exact solutions of the (0+1)-dimensional Boltzmann equation treated in the relaxation time approximation~\cite{Florkowski:2014sfa} by taking into account the effects of quantum statistics. As have been pointed out before~\cite{Florkowski:2013lza,Florkowski:2013lya}, such effects are trivial for the gas of massless particles. However, for the massive gas studied here, the effects of quantum statistics lead to interesting consequences --- they influence the form of the kinetic coefficients, and affect the time dynamics of the bulk viscous pressure.

The exact solutions of the Boltzmann equation are very useful for selection of the proper form of the kinetic coefficients~\cite{Florkowski:2013lza,Florkowski:2013lya} and also for the selection of the right structure of the hydrodynamic equations~\cite{Florkowski:2013lza,Florkowski:2013lya,Florkowski:2014bba,Nopoush:2014pfa,Denicol:2014xca,Denicol:2014tha,Nopoush:2014qba}. The latter usually represent the approximate description of the non-equilibrium systems that should be analyzed more thoroughly within the kinetic-theory framework.  In particular, the recent studies based on the exact solutions of the Boltzmann equations \cite{Denicol:2014mca,Jaiswal:2014isa} revealed the importance of the shear-bulk couplings in the hydrodynamic approach \cite{Denicol:2014vaa}. The proper description of the bulk pressure is important as it may affect different physical observables studied in relativistic heavy-ion collisions, as has been recently pointed out in Refs.~ \cite{Noronha-Hostler:2013gga,Noronha-Hostler:2013hsa,Rose:2014fba}. 

%%%%%%%%%%%%%%%%%%%%%%%%%%%%%%%%%%%%%%%%%%%%%%%%%%%%%%%%%%%%%%%%%%%%%%%%%%%%%%%%%%%%%%%%%%%%%%%%%%%%%
\section{The Boltzmann equation in relaxation time approximation}
\label{sect:kineq}
%%%%%%%%%%%%%%%%%%%%%%%%%%%%%%%%%%%%%%%%%%%%%%%%%%%%%%%%%%%%%%%%%%%%%%%%%%%%%%%%%%%%%%%%%%%%%%%%%%%%%

Our considerations are based on the relativistic Boltzmann equation 
\begin{equation}
 p^\mu \partial_\mu  f(x,p) =  C[f(x,p)] \, ,
\label{kineq}
\end{equation}
where $f(x,p)$ is the distribution function, and $C$ is the collision kernel treated in the relaxation time approximation (RTA)~\cite{1954PhRv...94..511B,Anderson:1974}
\begin{equation}
C[f] = - \frac{p_\mu u^\mu}{\tau_{\rm eq}} (f - f_{\rm eq}) .
\label{col-term}
\end{equation}
Here $\tau_{\rm eq}$ is the relaxation time, while $u^\mu$ is the flow velocity of matter. The (background) equilibrium distribution function $f_{\rm eq}$ has the form
\begin{equation}
f_{\rm eq} = \frac{2}{(2\pi)^3} \left[\exp\left(\frac{p_\mu u^\mu}{T} \right) - \epsilon \right]^{-1} ,
\label{Boltzmann}
\end{equation}
where the factor $\epsilon = 1$ ($\epsilon = -1$) corresponds to the Bose-Einstein (Fermi-Dirac) statistics. In the limit $\epsilon \to 0$ we reproduce an earlier studied case of the classical Boltzmann statistics~\cite{Florkowski:2014sfa}. We note that the factor of two in Eq.~(\ref{Boltzmann}) accounts for spin degeneracy.  The effective temperature $T$ appearing in (\ref{Boltzmann}) is obtained from the Landau matching which requires that the energy density obtained with the distribution function $f$ is the same as the energy density obtained with the equilibrium distribution $f_{\rm eq}$.  Only if the system is close to equilibrium,  the effective temperature $T$ can be interpreted as the genuine temperature of the system.  In more general cases $T$ should be treated as an alternative measure of the energy density. We also note that the use of Eqs.~(\ref{kineq})--(\ref{Boltzmann}) has been motivated by other studies, see e.g. Refs.~\cite{Anderson:1974,Czyz:1986mr,Dyrek:1986vv,cerc,Sasaki:2008fg,Bozek:2009dw,Romatschke:2011qp} where RTA was applied to calculate the transport properties of relativistic fluids.

%%%%%%%%%%%%%%%%%%%%%%%%%%%%%%%%%%%%%%%%%%%%%%%%%%%%%%%%%%%%%%%%%%%%%%%%%%%%%%%%%%%%%%%%%%%%%%%%%%%%%
\subsection{Equilibrium thermodynamic functions}
\label{sect:eqthermo}
%%%%%%%%%%%%%%%%%%%%%%%%%%%%%%%%%%%%%%%%%%%%%%%%%%%%%%%%%%%%%%%%%%%%%%%%%%%%%%%%%%%%%%%%%%%%%%%%%%%%%

For massive particles obeying Bose-Einstein or Fermi-Dirac statistics, the equilibrium particle density, entropy 
density, energy density, and pressure can be expressed as the series of the modified Bessel functions $K_n$~\cite{florkowski2010}, namely
\begin{eqnarray} 
{\cal N}_{\rm eq}(T) &=& \frac{g_0}{\pi^2}\, T m^2\, \epsilon\, \sum_{\kappa=1}^\infty \frac{\epsilon^\kappa}{\kappa} \, K_2\left( \frac{m}{T}\kappa\right), \label{neq} \\
{\cal S}_{\rm eq}(T) &=& \frac{g_0}{\pi^2} m^2\, \epsilon\, \sum_{\kappa=1}^\infty \frac{\epsilon^\kappa}{\kappa^2} \left[4\,T  K_2\left( \frac{m}{T}\kappa \right) + m\,\kappa\, K_1\left( \frac{m}{T}\kappa \right) \right] ,
\label{sigmaeq} \\
{\cal E}_{\rm eq}(T) &=& \frac{g_0 T m^2 }{\pi^2} 
\epsilon\, \sum_{\kappa=1}^\infty \frac{\epsilon^\kappa}{\kappa^2}
 \left[ 3T K_{2}\left( \frac{m}{T} \kappa \right) +m \kappa K_{1} \left( \frac{m}{T} \kappa \right) \right], 
\label{epsiloneq} \\
{\cal P}_{\rm eq}(T) &=& \frac{g_0}{\pi^2}\, T^2  m^2\, \epsilon\, \sum_{\kappa=1}^\infty \frac{\epsilon^\kappa}{\kappa^2} \, K_2\left( \frac{m}{T}\kappa\right),
\label{Peq}
\end{eqnarray}
where $g_0$ accounts for all internal degrees of freedom different from spin. We stress that the parameter $\epsilon$, which specifies the quantum statistics, will appear implicitly in the formalism. All physical observables are calculated for three possible options: $\epsilon = +1, -1, 0$. For brevity of notation, however, we do not use the label $\epsilon$ to distinguish between different versions of the functions appearing in the calculations.

Alternatively, one can show that the equilibrium energy density and pressure may be expressed by the equations
\begin{eqnarray} 
{\cal E}_{\rm eq}(T) &=& \frac{g_0 T^4}{2\pi^2} 
\tilde{\cal H}_2\left(1,\frac{m}{T}\right),
\label{epsiloneq2} \\
{\cal P}_{\rm eq}(T) &=& \frac{g_0}{3\pi^2}\, T^4 {\cal M}\left(\frac{m}{T}\right).
\label{Peq2}
\end{eqnarray}
Here the function $\tilde{\cal H}_2(y,z)$ is defined by the integral
\begin{eqnarray}
\tilde{\cal H}_2(y,z) &=& \int\limits_0^\infty du\, u^3 \, {\cal H}_2\left(y,\frac{z}{u} \right)
\, \left[ \exp\left(\sqrt{u^2+z^2}\right) - \epsilon \right]^{-1},
\label{tildeH2}
\end{eqnarray}
where $ {\cal H}_2(y,\zeta)$ is given by Eq.~(24) in \cite{Florkowski:2014sfa}. In the special case, where the first argument of $ {\cal H}_2$ is equal to unity,  as needed in Eq.~(\ref{epsiloneq2}), we have  ${\cal H}_2\left(1,\zeta\right) = 2\sqrt{1+\zeta^2}$.  The function ${\cal M}(z)$, needed to evaluate the equilibrium pressure in Eq.~(\ref{Peq2}),  is defined as the integral
\begin{eqnarray}
{\cal M}(z)=z^4\int\limits_0^{\infty}du \sinh^4 u \left[\exp(z \cosh u) - \epsilon \right]^{-1}.
\end{eqnarray}

%%%%%%%%%%%%%%%%%%%%%%%%%%%%%%%%%%%%%%%%%%%%%%%%%%%%%%%%%%%%%%%%%%%%%%%%%%%%%%%%%%%%%%%%%%%%%%%%%%%%%
\subsection{Boost-invariant variables}
\label{sect:boostinvvar}
%%%%%%%%%%%%%%%%%%%%%%%%%%%%%%%%%%%%%%%%%%%%%%%%%%%%%%%%%%%%%%%%%%%%%%%%%%%%%%%%%%%%%%%%%%%%%%%%%%%%%

Similarly to Ref.~\cite{Florkowski:2014sfa}, herein we consider the case of a transversely homogeneous boost-invariant system  --- all scalar functions of space and time  depend only on the proper time $\tau = \sqrt{t^2-z^2}$, while the flow has the Bjorken form, $u^\mu = \left(t/\tau,0,0,z/\tau\right)$~\cite{Bjorken:1982qr}.  The phase-space distribution functions transform also as scalars under Lorentz transformations, implying that $f(x,p)$ may depend only on $\tau$, $w$, and $\vec{p}_T$ with $w =  tp_L - z E$~\cite{Bialas:1984wv,Bialas:1987en}. Using $w$ and $p_L$ one can define another boost-invariant variable, namely
\begin{equation}
v(\tau,w,p_T) = Et-p_L z = 
\sqrt{w^2+\left( m^2+\vec{p}_T^{\,\,2}\right) \tau^2} \, .  
\label{v}
\end{equation}
The energy and longitudinal momentum of a particle is obtained from the inverse transformations: $E= p^0 = (vt+wz)/\tau^2$, and $p_L=(wt+vz)/\tau^2$. The boost-invariant measure in the momentum space is defined as
\begin{equation}
dP = 2 \, d^4p \, \delta \left( p^2-m^2\right) \theta (p^0)
= d^2p_T \frac{dp_L}{p^0} = d^2p_T\frac{dw}{v} \, .  
\label{dP}
\end{equation}
With the help of  the boost-invariant variables introduced above, we rewrite the kinetic equation (\ref{kineq}) as
\begin{equation}
\frac{\partial f}{\partial \tau} = 
\frac{f_{\rm eq}-f}{\tau_{\rm eq}},
\label{kineqs}
\end{equation}
where the equilibrium function  (\ref{Boltzmann}) takes now the following quantum-statistics dependent form
\begin{equation}
f_{\rm eq}(\tau, w, p_T) =
\frac{2}{(2\pi)^3} \left\{ \exp\left[
\frac{\sqrt{w^2+ \left( m^2+p_T^2 \right) \tau^2}}{T(\tau) \tau} \,  \right] -\epsilon\right\}^{-1}.
\label{feq}
\end{equation}
Similarly to our previous studies, we assume that $f(\tau,w,\vec{p}_T)$  is an even function of $w$, and depends only on  $|\vec{p}_T|$.

%%%%%%%%%%%%%%%%%%%%%%%%%%%%%%%%%%%%%%%%%%%%%%%%%%%%%%%%%%%%%%%%%%%%%%%%%%%%%%%%%%%%%%%%%%%%%%%%%%%%%
\subsection{Energy-momentum conservation}
\label{sect:enmomcom}
%%%%%%%%%%%%%%%%%%%%%%%%%%%%%%%%%%%%%%%%%%%%%%%%%%%%%%%%%%%%%%%%%%%%%%%%%%%%%%%%%%%%%%%%%%%%%%%%%%%%%

The energy-momentum tensor is the second moment of the phase-space distribution function. It can be written in the form \cite{Florkowski:2011jg,Martinez:2012tu}
\begin{equation}
T^{\mu\nu} = ({\cal E} + {\cal P}_T) u^\mu u^\nu - {\cal P}_T g^{\mu\nu} + ({\cal P}_L-{\cal P}_T) z^\mu z^\nu 
\, ,
\label{Tmunu2}
\end{equation}
where the energy density ${\cal E}$, the longitudinal pressure ${\cal P}_L$, and the transverse pressure ${\cal P}_T$, are obtained from the integrals
\begin{eqnarray} 
{\cal E}(\tau) &= \frac{g_0}{\tau^2}\,
\int dP \, v^2\,  f(\tau,w,p_T) \, , \\
{\cal P}_L(\tau) &= \frac{g_0}{\tau^2}\,
\int dP \, w^2\,  f(\tau,w,p_T)\, \\
{\cal P}_T(\tau) &= \frac{g_0}{2}\,
\int dP \, p_T^2\, f(\tau,w,p_T) \, ,
\label{epsandpres}
\end{eqnarray}
and $z^\mu = \left(z/\tau,0,0,t/\tau\right)$. In our case, the energy-momentum conservation law requires that $d{\cal E}/{d\tau}=
- ({\cal E}+{\cal P}_L)/{\tau}$.  We note that the energy-momentum tensor (\ref{Tmunu2}) has the structure typical for a~momentum-space anisotropic system \cite{Florkowski:2011jg,Martinez:2012tu}. The energy conservation equation is satisfied if the energy densities calculated with  $f$ and  $f_{\rm eq}$ are equal, which leads to the Landau matching condition
\begin{eqnarray}
\hspace{-1.75cm} {\cal E}(\tau) &=& \frac{g_0}{\tau^2} \,
\int dP \, v^2\,  f(\tau,w,p_T) = \frac{g_0}{\tau^2}\,
\int dP \, v^2\,  f_{\rm eq}(\tau,w,p_T) =   \frac{g_0 T^4}{2\pi^2} 
\tilde{\cal H}_2\left(1,\frac{m}{T}\right).  \label{LM1}
\end{eqnarray}

%%%%%%%%%%%%%%%%%%%%%%%%%%%%%%%%%%%%%%%%%%%%%%%%%%%%%%%%%%%%%%%%%%%%%%%%%%%%%%%%%%%%%%%%%%%%%%%%%%%%%
\section{Solutions of kinetic equation}
\label{sect:formsol}
%%%%%%%%%%%%%%%%%%%%%%%%%%%%%%%%%%%%%%%%%%%%%%%%%%%%%%%%%%%%%%%%%%%%%%%%%%%%%%%%%%%%%%%%%%%%%%%%%%%%%

%%%%%%%%%%%%%%%%%%%%%%%%%%%%%%%%%%%%%%%%%%%%%%%%%%%%%%%%%%%%%%%%%%%%%%%%%%%%%%%%%%%%%%%%%%%%%%%%%%%%%
\subsection{General form of solutions}
\label{sect:genformsol}
%%%%%%%%%%%%%%%%%%%%%%%%%%%%%%%%%%%%%%%%%%%%%%%%%%%%%%%%%%%%%%%%%%%%%%%%%%%%%%%%%%%%%%%%%%%%%%%%%%%%%

The general solutions of Eq.~(\ref{kineqs}) have been found in Refs.~\cite{Florkowski:2013lza,Florkowski:2013lya,
Baym:1984np,Baym:1985tna,Heiselberg:1995sh,Wong:1996va} and read
\begin{equation}
f(\tau,w,p_T) = D(\tau,\tau_0) f_0(w,p_T)  + \int_{\tau_0}^\tau \frac{d\tau^\prime}{\tau_{\rm eq}(\tau^\prime)} \, D(\tau,\tau^\prime) \, 
f_{\rm eq}(\tau^\prime,w,p_T) \, ,  \label{solf}
\end{equation}
where $D(\tau_2,\tau_1)$ is the damping function
\begin{equation}
D(\tau_2,\tau_1) = \exp\left[-\int_{\tau_1}^{\tau_2}
\frac{d\tau^{\prime\prime}}{\tau_{\rm eq}(\tau^{\prime\prime})} \right].
\end{equation}
In this paper, we use the initial conditions set at $\tau=\tau_0$. The initial distribution function $f$ is introduced in Romatschke-Strickland (RS) form with an underlying Bose-Einstein or Fermi-Dirac distribution as the isotropic distribution~\cite{Romatschke:2003ms}
\begin{eqnarray}
f_0(w,p_T) &=& \frac{2}{(2\pi)^3}
\left\{ \exp\left[
\frac{\sqrt{(p\cdot u)^2 + \xi_0 (p\cdot z)^2}}{\Lambda_0} \, \right] - \epsilon \right\}^{-1 } \nonumber \\
&=& \frac{1}{4\pi^3}
\left\{ \exp\left[
\frac{\sqrt{(1+\xi_0) w^2 + (m^2+p_T^2) \tau_0^2}}{\Lambda_0 \tau_0}\, \right] -\epsilon \right\}^{-1}.
\label{f0}
\end{eqnarray}
This form simplifies to an equilibrium Bose-Einstein or Fermi-Dirac distribution if the anisotropy parameter $\xi_0$ is zero, in which case the transverse momentum scale $\Lambda_0$ can be identified with the system's initial temperature $T_0$. 

%%%%%%%%%%%%%%%%%%%%%%%%%%%%%%%%%%%%%%%%%%%%%%%%%%%%%%%%%%%
\subsection{Dynamical Landau matching}
\label{sect:LM}
%%%%%%%%%%%%%%%%%%%%%%%%%%%%%%%%%%%%%%%%%%%%%%%%%%%%%%%%%%%

We multiply Eqs.~(\ref{feq}) and (\ref{f0}) by $g_0 v^2/\tau^2$, and integrate over momentum to get
\begin{eqnarray}
\frac{g_0}{\tau^2}\,\int dP \, v^2\,  f_{\rm eq}(\tau^\prime,w,p_T)
&=& \frac{g_0 T^4(\tau^\prime)}{2\pi^2} 
\, \tilde{\cal H}_2\left[ \frac{\tau^\prime}{\tau},\frac{m}{T(\tau^\prime)}\right],
\label{intfeq} \\
\frac{g_0}{\tau^2}\, \int dP \, v^2\,  f_0(w,p_T) 
&=& \frac{g_0 \Lambda^4_0}{2\pi^2} 
\, \tilde{\cal H}_2\left[ \frac{\tau_0}{\tau \sqrt{1+\xi_0}},\frac{m}{\Lambda_0}\right]. \label{intf0}
\end{eqnarray}
In the next step, from  Eqs.~(\ref{LM1}), (\ref{solf}), (\ref{intfeq}), and (\ref{intf0}) we obtain our main equation
\begin{eqnarray}
&&  
T^4(\tau)  \tilde{\cal H}_2\left[1,\frac{m}{T(\tau)}\right]
\label{LM2}  \\
&& 
 = D(\tau,\tau_0) \Lambda^4_0 \tilde{\cal H}_2\left[ \frac{\tau_0}{\tau \sqrt{1+\xi_0}},\frac{m}{\Lambda_0}\right]  + \int\limits_{\tau_0}^\tau 
\frac{d\tau^\prime}{\tau_{\rm eq}} D(\tau,\tau^\prime)
T^4(\tau^\prime) 
\tilde{\cal H}_2\left[ \frac{\tau^\prime}{\tau},\frac{m}{T(\tau^\prime)}\right]. \nonumber
\end{eqnarray}
The integral equation for  $T(\tau)$ is solved with the iterative method \cite{Banerjee:1989by}. We note that the form of Eqs.~(\ref{intfeq})--(\ref{LM2}) is the same as that found previously in the case of the Boltzmann statistics. The only difference resides in the implicit dependence of the functions $ f_{\rm eq}$, $f_0$, and $\tilde{\cal H}_2$ on the quantum statistics parameter $\epsilon$.

%%%%%%%%%%%%%%%%%%%%%%%%%%%%%%%%%%%%%%%%%%%%%%%%%%%%%%%%%%%%%%%%%%%%%%%%%%%%%%%%%%%%%%%%%%%%%%%%%%%%
\subsection{Transverse and longitudinal pressures}
\label{sect:tandlpres}
%%%%%%%%%%%%%%%%%%%%%%%%%%%%%%%%%%%%%%%%%%%%%%%%%%%%%%%%%%%%%%%%%%%%%%%%%%%%%%%%%%%%%%%%%%%%%%%%%%%%%

The degree of a system's equilibration can be obtained by comparing its transverse and longitudinal pressures.  The transverse pressure is obtained from the formula
\begin{eqnarray}
{\cal P}_T(\tau) &=& \frac{g_0}{4\pi^2} D(\tau,\tau_0)
\Lambda_0^4 \tilde{\cal H}_{2T}\left[ \frac{\tau_0}{\tau \sqrt{1+\xi_0}},\frac{m}{\Lambda_0}\right]
\nonumber \\ && 
+ \frac{g_0 }{4 \pi^2} \int\limits_{\tau_0}^\tau 
\frac{d\tau^\prime}{ \tau_{\rm eq}} 
 D(\tau,\tau^\prime)
T^4(\tau^\prime) 
\tilde{\cal H}_{2T}\left[ \frac{\tau^\prime}{\tau},\frac{m}{T(\tau^\prime)}\right],
\label{PT}
\end{eqnarray}
where 
\begin{equation}
\tilde{\cal H}_{2T}(y,z) = \int\limits_0^\infty du\, u^3 \, {\cal H}_{2T}
\left(y,\frac{z}{u} \right) \, \left[ \exp\left(\sqrt{u^2+z^2}\right) - \epsilon \right]^{-1}
\label{tildeH2T}
\end{equation}
with the function ${\cal H}_{2T}(y,\zeta)$ defined by Eq.~(30) in Ref.~\cite{Florkowski:2014sfa}. Similarly, the longitudinal pressure is obtained from
\begin{eqnarray}
{\cal P}_L(\tau) &=& \frac{g_0}{2\pi^2} D(\tau,\tau_0)
\Lambda_0^4 \tilde{\cal H}_{2L}\left[ \frac{\tau_0}{\tau \sqrt{1+\xi_0}},\frac{m}{\Lambda_0}\right]
\nonumber \\
&& + \frac{g_0 }{2 \pi^2} \int\limits_{\tau_0}^\tau 
\frac{d\tau^\prime}{ \tau_{\rm eq}}  D(\tau,\tau^\prime)
T^4(\tau^\prime) 
\tilde{\cal H}_{2L}\left[ \frac{\tau^\prime}{\tau},\frac{m}{T(\tau^\prime)}\right], 
\label{PL}
\end{eqnarray}
where the function $\tilde{\cal H}_{2L}$ is defined by the integral
\begin{equation}
\tilde{\cal H}_{2L}(y,z) = \int\limits_0^\infty du\, u^3 \, {\cal H}_{2L}
\left(y,\frac{z}{u} \right)
\, \left[ \exp\left(\sqrt{u^2+z^2}\right) - \epsilon \right]^{-1},
\label{tildeH2L}
\end{equation}
and the function ${\cal H}_{2L}(y,\zeta)$ is defined by Eq.~(35) in Ref.~\cite{Florkowski:2014sfa}.

%%%%%%%%%%%%%%%%%%%%%%%%%%%%%%%%%%%%%%%%%%%%%%%%%%%%%%%%%%%%%%%%%%%%%%%%%%%%%%%%%%%%%%%%%%%%%%%%%%%%
\section{Shear and bulk viscosities of a relativistic quantum massive gas}
\label{sect:shearbulk}
%%%%%%%%%%%%%%%%%%%%%%%%%%%%%%%%%%%%%%%%%%%%%%%%%%%%%%%%%%%%%%%%%%%%%%%%%%%%%%%%%%%%%%%%%%%%%%%%%%%%%

To find the shear viscosity $\eta$ for the Bose--Einstein and Fermi--Dirac gases, we use the integral formula (see, for example, Ref.~\cite{Sasaki:2008fg})
\begin{eqnarray}
\eta(T)=\frac{2g_0\tau_{\rm eq}}{15T}\int\frac{d^3p}{(2\pi)^3}\frac{p^4}{E^2}\, f_{\rm eq}(1 +\epsilon f_{\rm eq}).
\label{eta_qs}
\end{eqnarray}
Using the exact solution of the Boltzmann equation, we determine the effective shear viscosity from the equation
\begin{equation}
\eta_{\rm eff}(\tau) = \frac{1}{2} \, \tau \, \left[{\cal P}_T(\tau)-{\cal P}_L(\tau)\right]. 
\label{etakin}
\end{equation}
The form (\ref{etakin}) follows from the structure of the energy-momentum tensor in boost-invariant first-order viscous hydrodynamics. Therefore, one expects that the results obtained using Eqs.~(\ref{eta_qs}) and (\ref{etakin}) will agree only at late times, i.e., for $\tau \gg \tau_{\rm eq}$, when the system approaches equilibrium. We note, that to make a comparison between Eqs.~(\ref{eta_qs}) and (\ref{etakin}) one should include the time dependence of the effective temperature $T$ in Eq.~(\ref{eta_qs}), which follows directly from the Landau matching.

In the similar way we treat the bulk viscosity $\zeta$. For a quantum massive gas, the formula for the bulk viscosity is (see again Ref.~\cite{Sasaki:2008fg})
\begin{eqnarray}
\zeta(T)=\frac{2g_0\tau_{\rm eq}}{3T}\int\frac{d^3p}{(2\pi)^3}\frac{m^2}{E^2}
f_{\rm eq}(1 + \epsilon f_{\rm eq})\left(c_s^2E-\frac{p^2}{3E}  \right).
\label{zeta_qs}
\end{eqnarray}
The effective bulk viscosity is obtained from the exact solution as
\begin{equation}
\zeta_{\rm eff}(\tau) = -\frac{1}{3} \tau
\left[{\cal P}_L(\tau) + 2 {\cal P}_T(\tau)
- 3 {\cal P}_{\rm eq}(\tau) \right] .
\label{zetakin}
\end{equation}
The sound velocity appearing in Eq.~(\ref{zeta_qs}) is obtained from the formula $c_s^2(T) = \partial {\cal P}_{\rm eq}(T)/\partial {\cal E}_{\rm eq}(T)$, where the equilibrium thermodynamic functions are given by Eqs.~(\ref{epsiloneq}) and (\ref{Peq}), or Eqs.~(\ref{epsiloneq2}) and (\ref{Peq2}). In the classical limit, $\epsilon \to 0$, the integrals (\ref{eta_qs}) and (\ref{zeta_qs}) become analytic. The appropriate formulas can be found in~Ref.~\cite{Florkowski:2014sfa}.

%%%%%%%%%%%%%%%%%%%%%%%%%%%%%%%%%%%%%%%%%%%%%%%%%%%%%%%%%%%%%%%%%%%%%%%%%%%%%%%%%%%%%%%%%%%%%%%%%%%%
\section{Results}
\label{sect:results}
%%%%%%%%%%%%%%%%%%%%%%%%%%%%%%%%%%%%%%%%%%%%%%%%%%%%%%%%%%%%%%%%%%%%%%%%%%%%%%%%%%%%%%%%%%%%%%%%%%%%%

%
\begin{figure}[t]
\begin{center}
\begin{minipage}[b]{7.25cm}
\centering
\includegraphics[angle=0,width=0.9\textwidth]{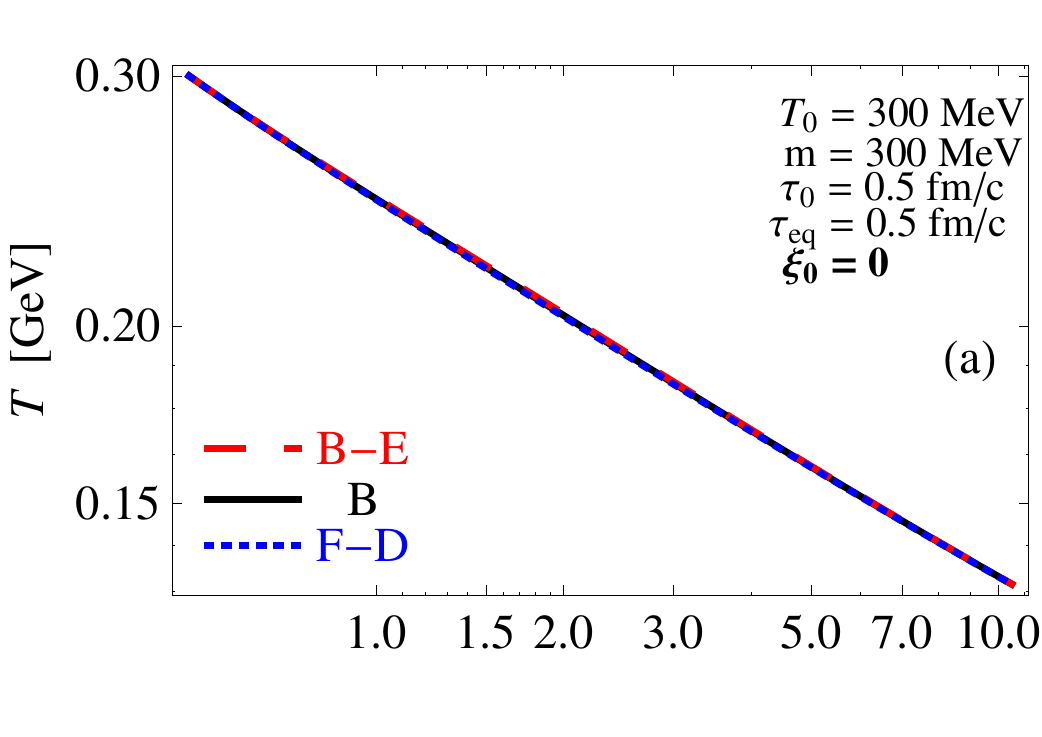}
\end{minipage}
\vspace{-0.5cm}
\begin{minipage}[b]{7.25cm}
\centering
\includegraphics[angle=0,width=0.9\textwidth]{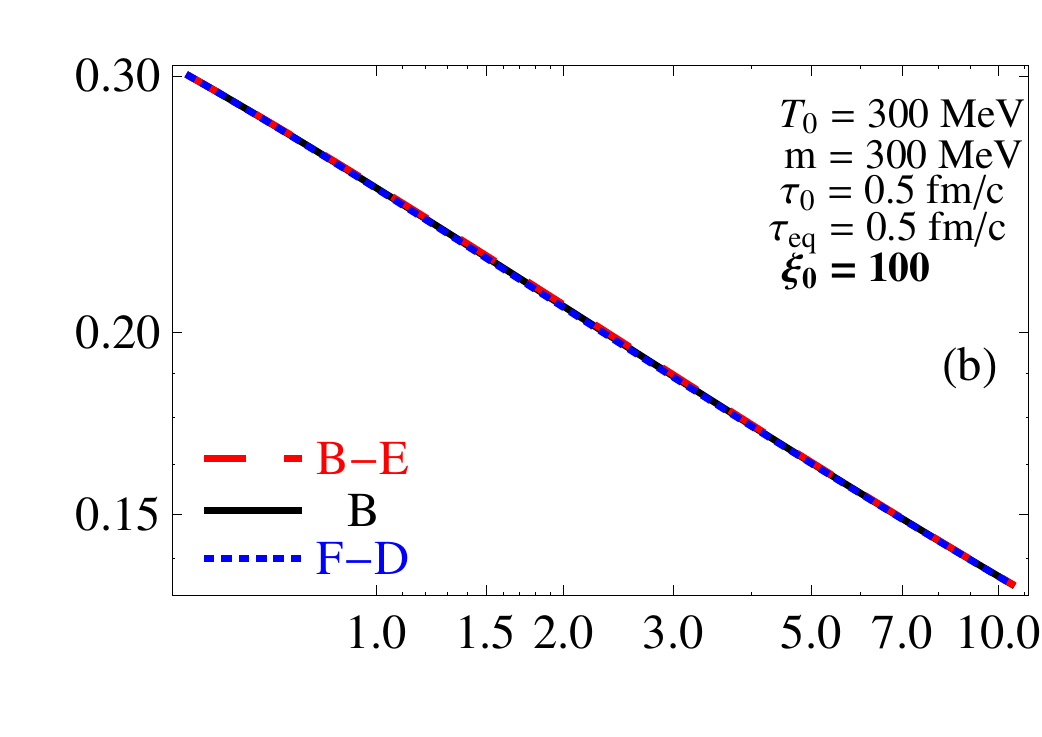}
\end{minipage}
\begin{minipage}[b]{7.25cm}
\centering
\includegraphics[angle=0,width=0.9\textwidth]{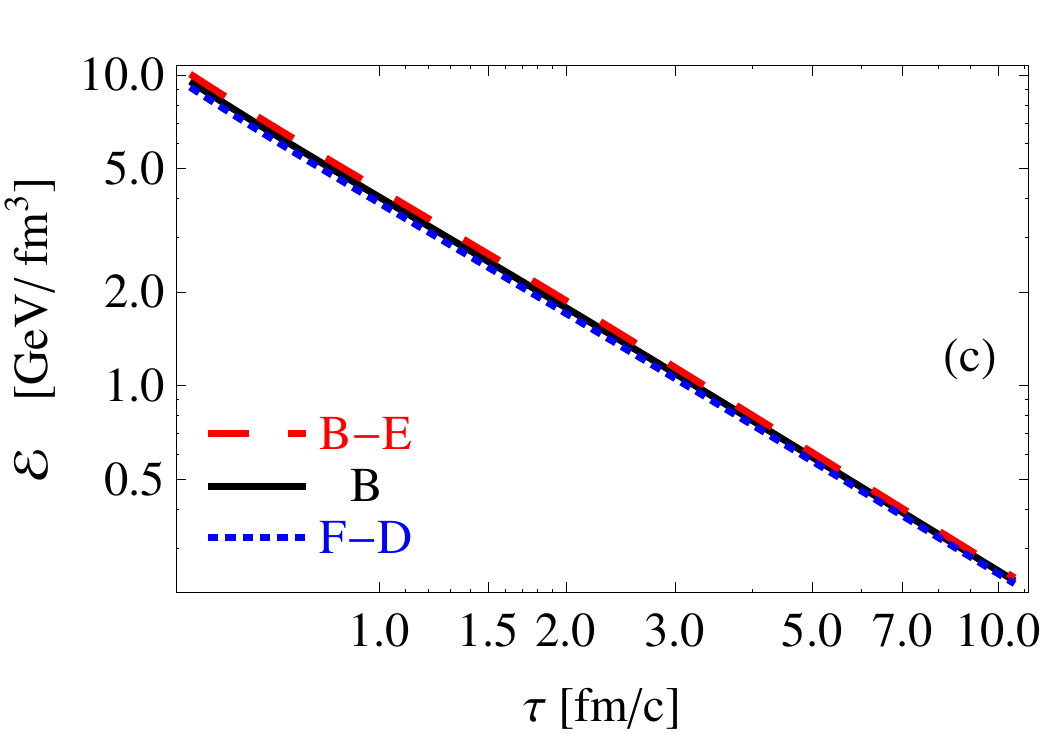}
\end{minipage}
\begin{minipage}[b]{7.25cm}
\centering
\includegraphics[angle=0,width=0.9\textwidth]{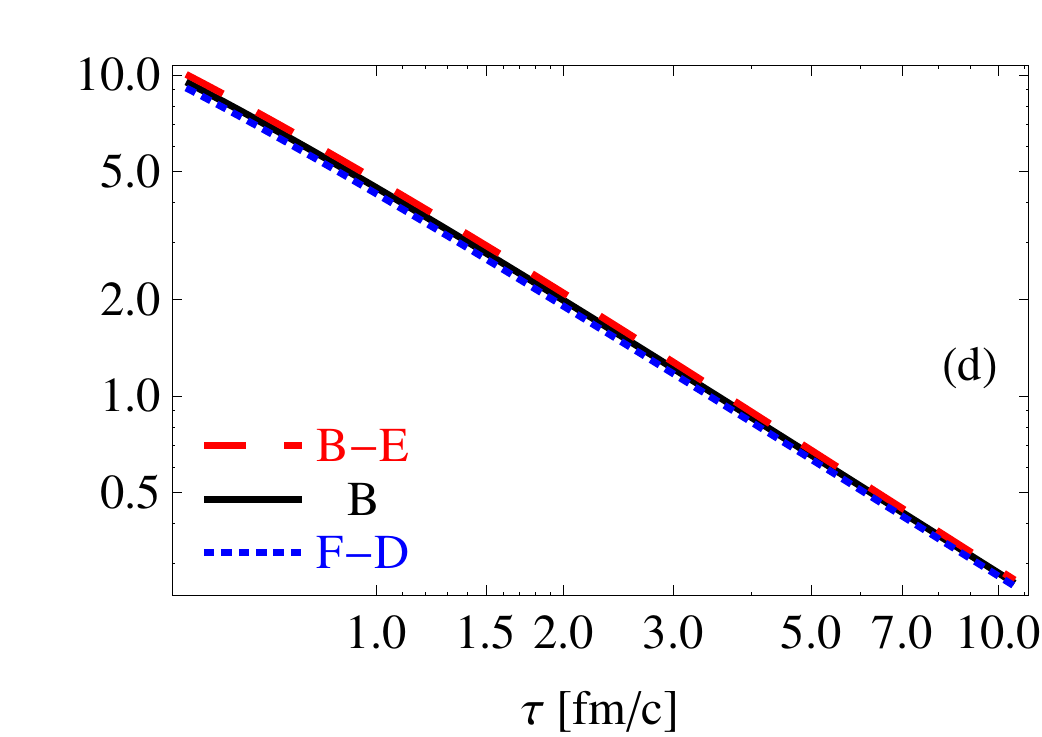}
\end{minipage}
\caption{(Color online) Effective temperature (two upper panels) and energy density (two lower panels) vs. proper time $\tau$, obtained with the initial condition $T_0=300\rm \,\,MeV$ at $\tau_0=0.5~\rm$~fm/c, and with the equilibration time  $\tau_{\rm eq}=0.5~\rm$~fm/c. In the cases (a) and (c) the initial system is isotropic ($\xi_0=0$), while in the cases (b) and (d) the initial system is highly oblate  ($\xi_0=100$). The mass used in the calculations is $m=$~300~MeV. The dashed, solid, and dotted lines describe the results obtained for the Bose-Einstein (B--E), Boltzmann (B) and Fermi-Dirac (F--D) statistics, respectively.}
\label{fig:Ene_Temp_T0-300}
\end{center}
\end{figure}
%

%%%%%%%%%%%%%%%%%%%%%%%%%%%%%%%%%%%%%%%%%%%%%%%%%%%%%%%%%%%%%%%%%%%%%%%%%%%%%%%%%%%%%%%%%%%%%%%%%%%%
\subsection{Initial conditions}
\label{sect:initcond}
%%%%%%%%%%%%%%%%%%%%%%%%%%%%%%%%%%%%%%%%%%%%%%%%%%%%%%%%%%%%%%%%%%%%%%%%%%%%%%%%%%%%%%%%%%%%%%%%%%%%%

In this Section we present our results obtained for different quantum statistics. We perform our numerical calculations for the fixed initial effective temperature, $T_0$~=~300~MeV. The equilibration time $\tau_{\rm eq}$ is kept constant, and equal to 0.5 fm/c.  The integral equation (\ref{LM2}) is solved by the iterative method~\cite{Banerjee:1989by}.  The initial time is taken to be \mbox{$\tau_0$ = 0.5 fm/c}, and we continue the evolution until \mbox{$\tau$ = 10.5 fm/c}. The degeneracy factor $g_0$ is taken to be 16, however, its specific value is irrelevant for our conclusions, since it either cancels in ratios we consider or appears as an overall factor. The initial distribution function is assumed in the RS form (\ref{f0}) with the initial anisotropy parameter $\xi_0 \in \{0,100\}$, corresponding to an initially isotropic or highly oblate initial configuration, respectively. The transverse-momentum scale $\Lambda_0$ is chosen in such a way that the initial energy density of an anisotropic system coincides with the energy density of an equilibrium system with temperature $T_0$
\begin{equation}
T_0^4
\, \tilde{\cal H}_2\left[1,\frac{m}{T_0}\right]
 =  \Lambda^4_0
\, \tilde{\cal H}_2\left[ \frac{1}{ \sqrt{1+\xi_0}},\frac{m}{\Lambda_0}\right],
\label{initL0}
\end{equation} 
which is simply the Landau matching condition (\ref{LM2}) at $\tau=\tau_0$. We note that for fixed $T_0$ and $\xi_0$, the value of $\Lambda_0$ depends on $m$ and, implicitly, on the statistics parameter $\epsilon$. In the present calculations we take $m=300$~MeV.

%----------------------------------------------------------------------------------------------------

%%%%%%%%%%%%%%%%%%%%%%%%%%%%%%%%%%%%%%%%%%%%%%%%%%%%%%%%%%%%%%%%%%%%%%%%%%%%%%%%%%%%%%%%%%%%%%%%%%%%
\subsection{Thermodynamics-like variables}
\label{sect:efftemp}
%%%%%%%%%%%%%%%%%%%%%%%%%%%%%%%%%%%%%%%%%%%%%%%%%%%%%%%%%%%%%%%%%%%%%%%%%%%%%%%%%%%%%%%%%%%%%%%%%%%%%

In Fig.~\ref{fig:Ene_Temp_T0-300} we plot the effective temperature $T$, panels (a) and (b), and the corresponding energy density, panels (c) and (d), both as functions of the proper time $\tau$. In the left column, panels (a) and (c), we show the results for the case where the initial system is isotropic, while in the right column, panels (b) and (d),  the initial system is highly oblate.  The dashed, solid, and dotted lines describe the results obtained for the Bose-Einstein (B--E), Boltzmann (B), and Fermi-Dirac (F--D) statistics, respectively. The results shown in Fig.~\ref{fig:Ene_Temp_T0-300}  suggest that the inclusion of quantum statistics has little effect on the time dependence of the effective temperature and the energy density.

\begin{figure}[t]
\begin{center}
\begin{minipage}[b]{7.25cm}
\centering
\includegraphics[angle=0,width=0.9\textwidth]{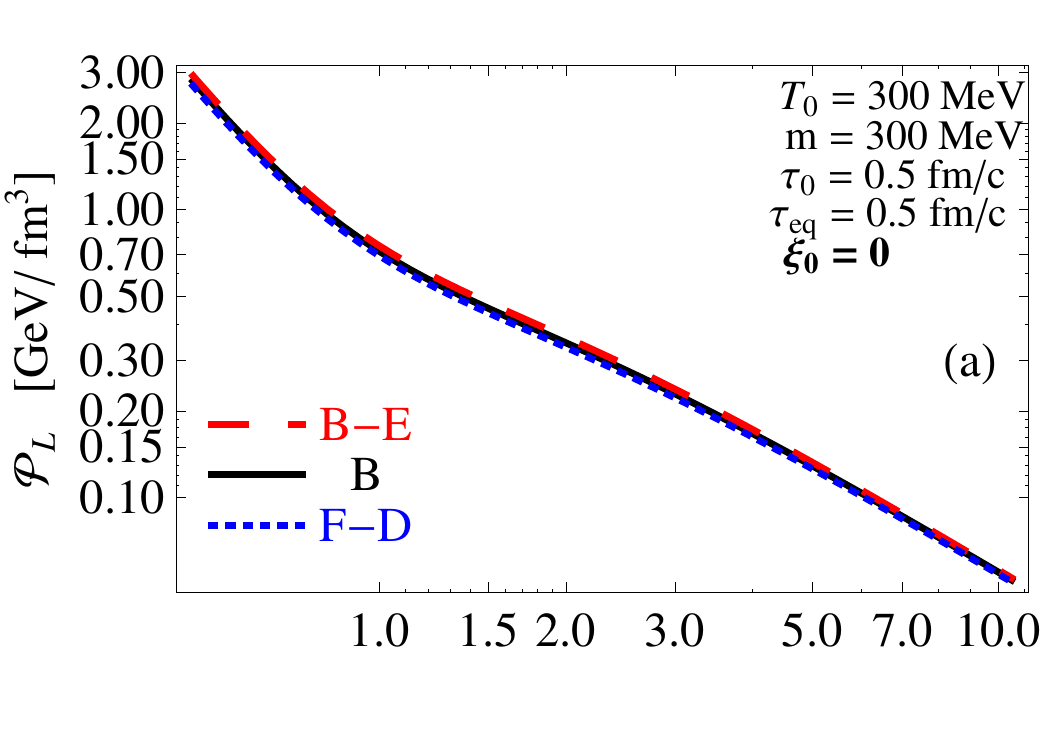}
\end{minipage}
\vspace{-0.5cm}
\begin{minipage}[b]{7.25cm}
\centering
\includegraphics[angle=0,width=0.9\textwidth]{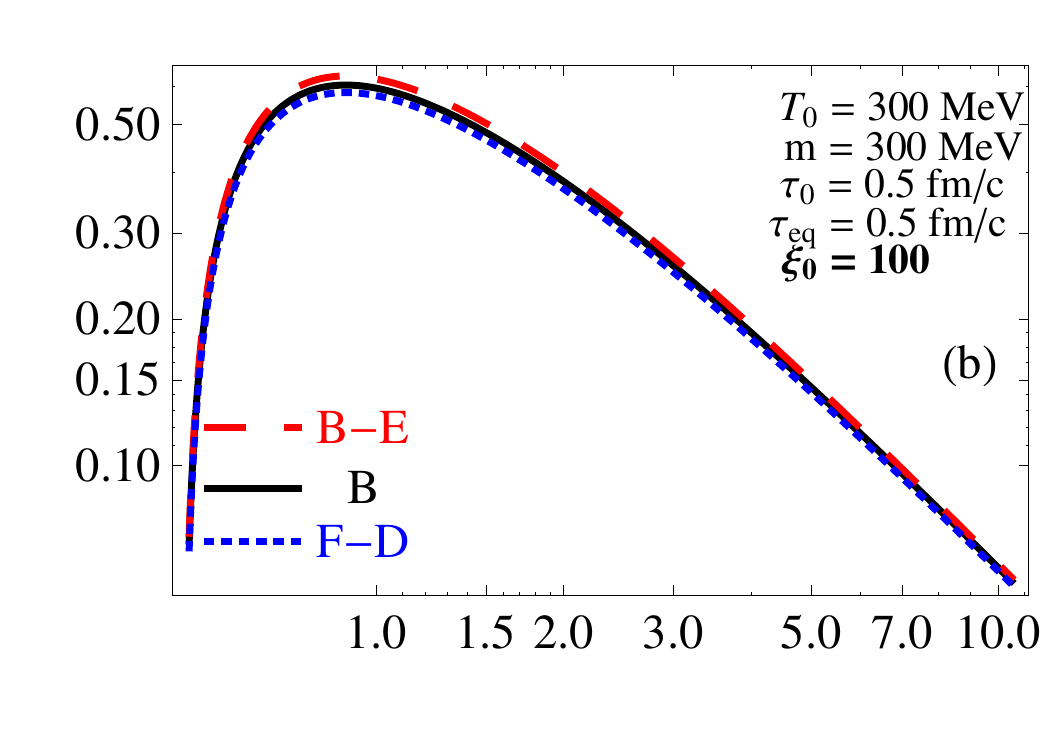}
\end{minipage}
\begin{minipage}[b]{7.25cm}
\centering
\includegraphics[angle=0,width=0.9\textwidth]{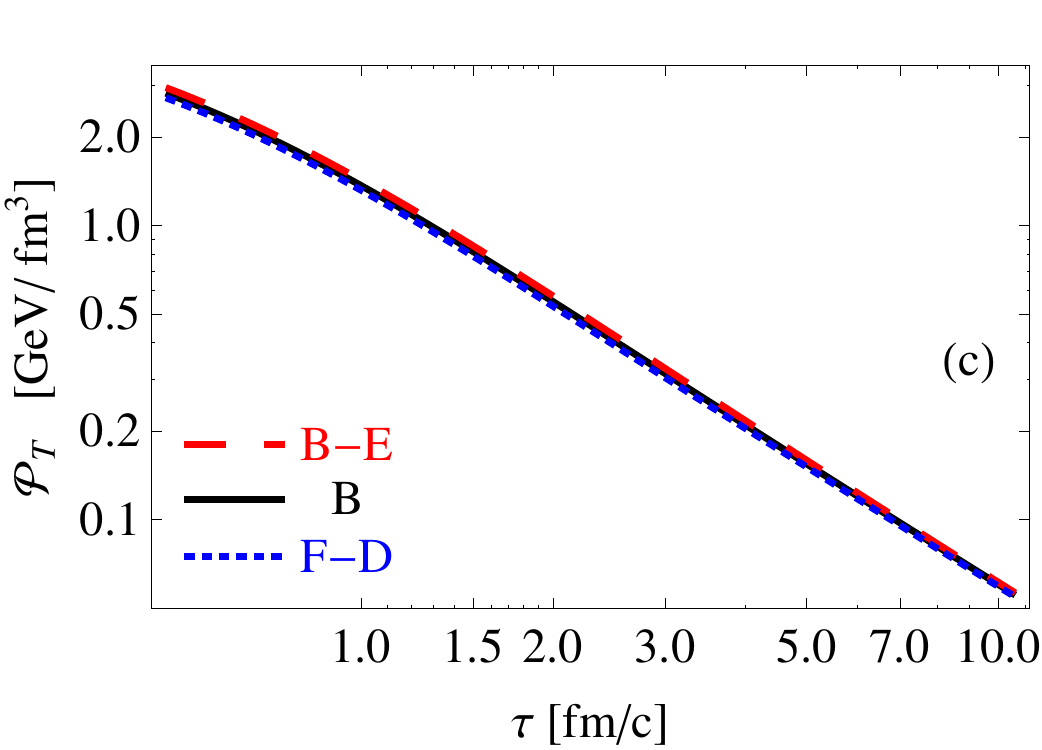}
\end{minipage}
\begin{minipage}[b]{7.25cm}
\centering
\includegraphics[angle=0,width=0.9\textwidth]{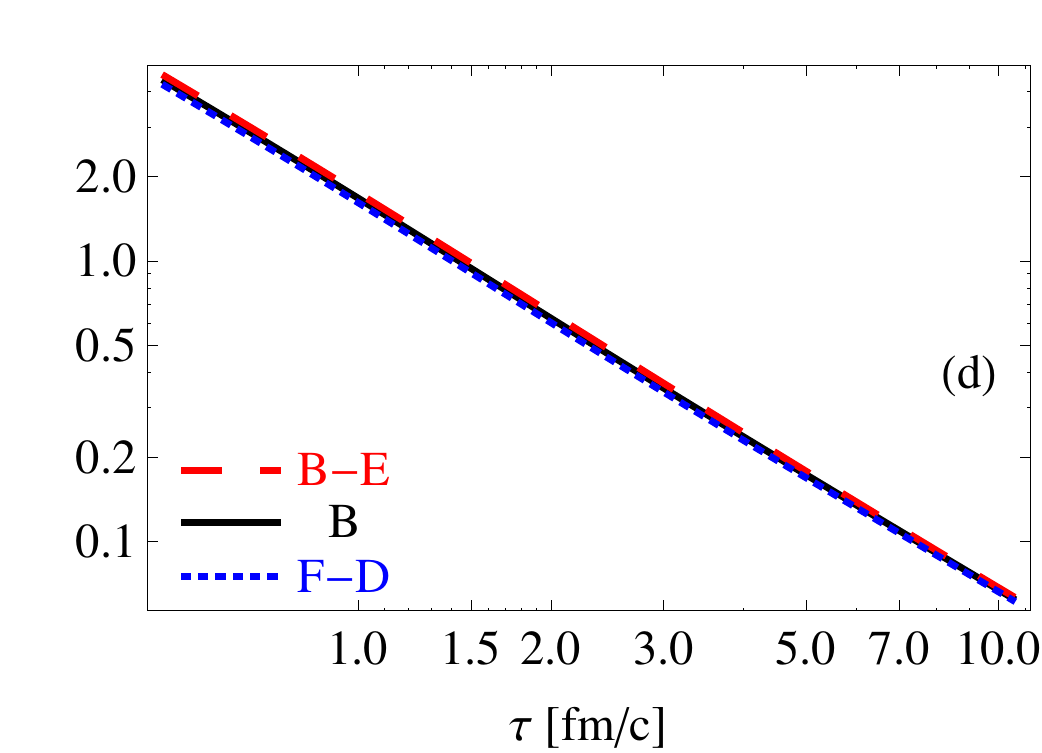}
\end{minipage}
\caption{(Color online) Thermodynamics-like variables: longitudinal and transverse pressure for $T_0=300\rm \,\,MeV$, $\tau_{\rm eq}=0.5 \rm $ fm/c. Plots (a) and (c) correspond to the initial isotropic system, (b) and (d) are for the initial anisotropic system.}
\label{fig:PL_PT_T0-300}
\end{center}
\end{figure}

In Fig.~\ref{fig:PL_PT_T0-300} we plot the longitudinal pressure, panels (a) and (b), and the transverse pressure, panels (c) and (d), as functions of the proper time. Again, the left column describes the case where the initial system is isotropic, while the right column describes the case where the initial system is oblate. Similarly as in Fig.~\ref{fig:Ene_Temp_T0-300}, we observe little dependence of the two components of pressure on the quantum statistics of particles.

%%%%%%%%%%%%%%%%%%%%%%%%%%%%%%%%%%%%%%%%%%%%%%%%%%%%%%%%%%%%%%%%%%%%%%%%%%%%%%%%%%%%%%%%%%%%%%%%%%%%
\subsection{Shear and bulk viscosity}
\label{sect:shearres}
%%%%%%%%%%%%%%%%%%%%%%%%%%%%%%%%%%%%%%%%%%%%%%%%%%%%%%%%%%%%%%%%%%%%%%%%%%%%%%%%%%%%%%%%%%%%%%%%%%%%%

\begin{figure}[t]
\begin{center}
\begin{minipage}[b]{5.05cm}
\centering
\includegraphics[angle=0,width=1\textwidth]{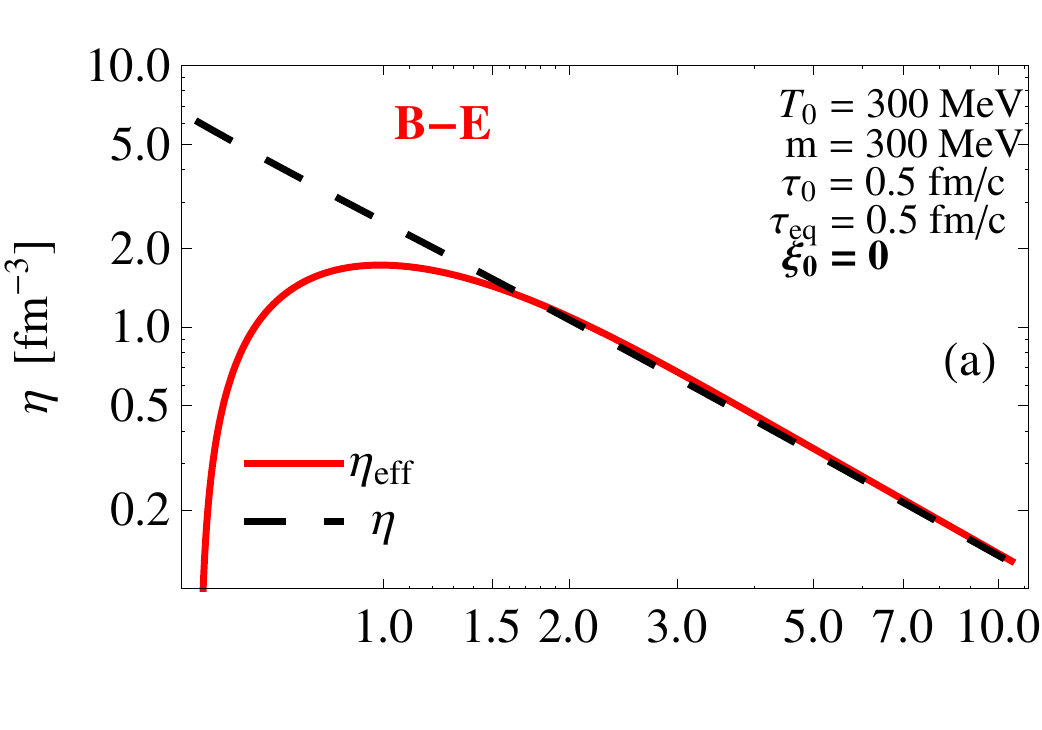}
\end{minipage}
\vspace{-0.35cm}
\begin{minipage}[b]{5.05cm}
\centering
\includegraphics[angle=0,width=1\textwidth]{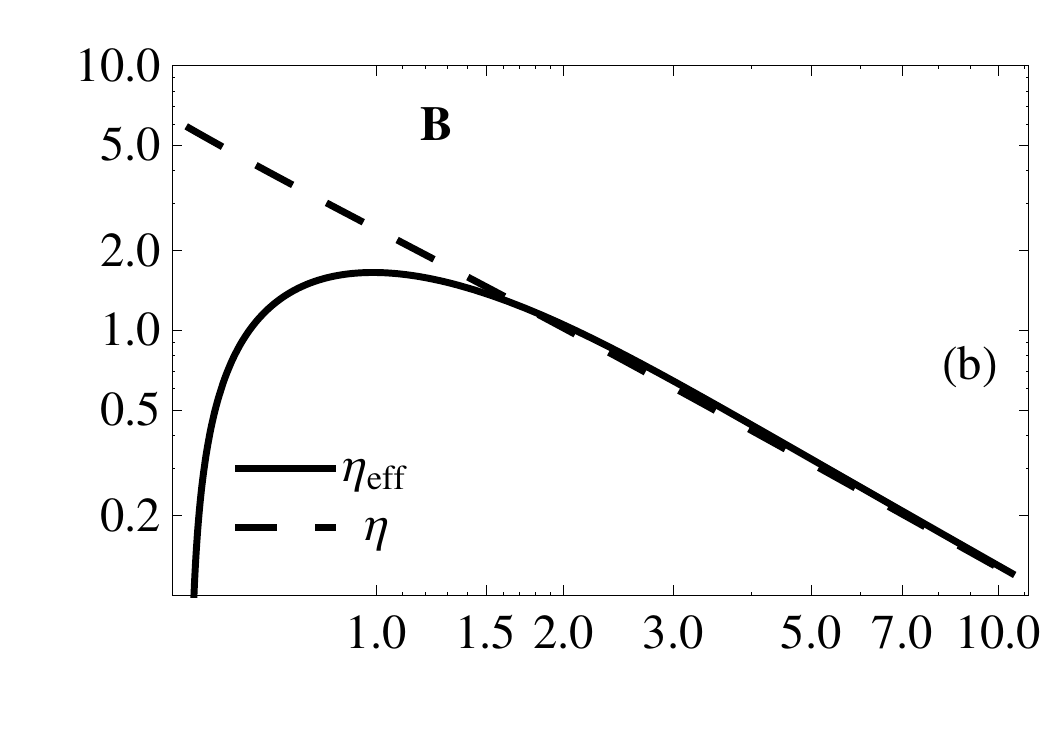}
\end{minipage}
\begin{minipage}[b]{5.05cm}
\centering
\includegraphics[angle=0,width=1\textwidth]{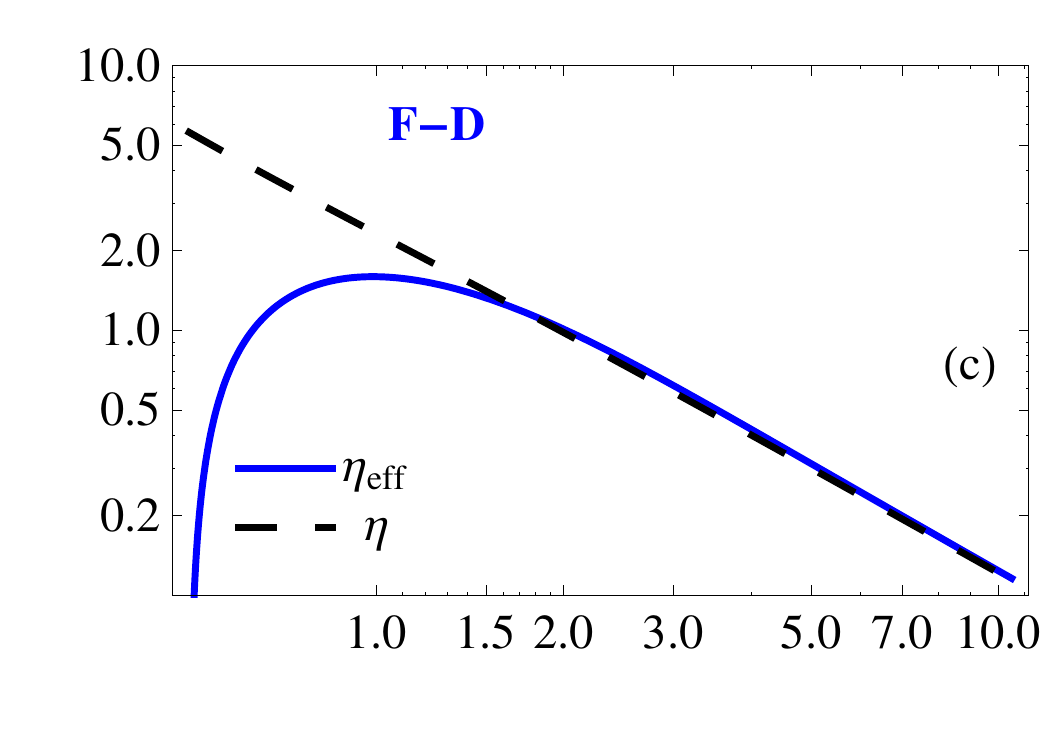}
\end{minipage}
\begin{minipage}[b]{5.05cm}
\centering
\includegraphics[angle=0,width=1\textwidth]{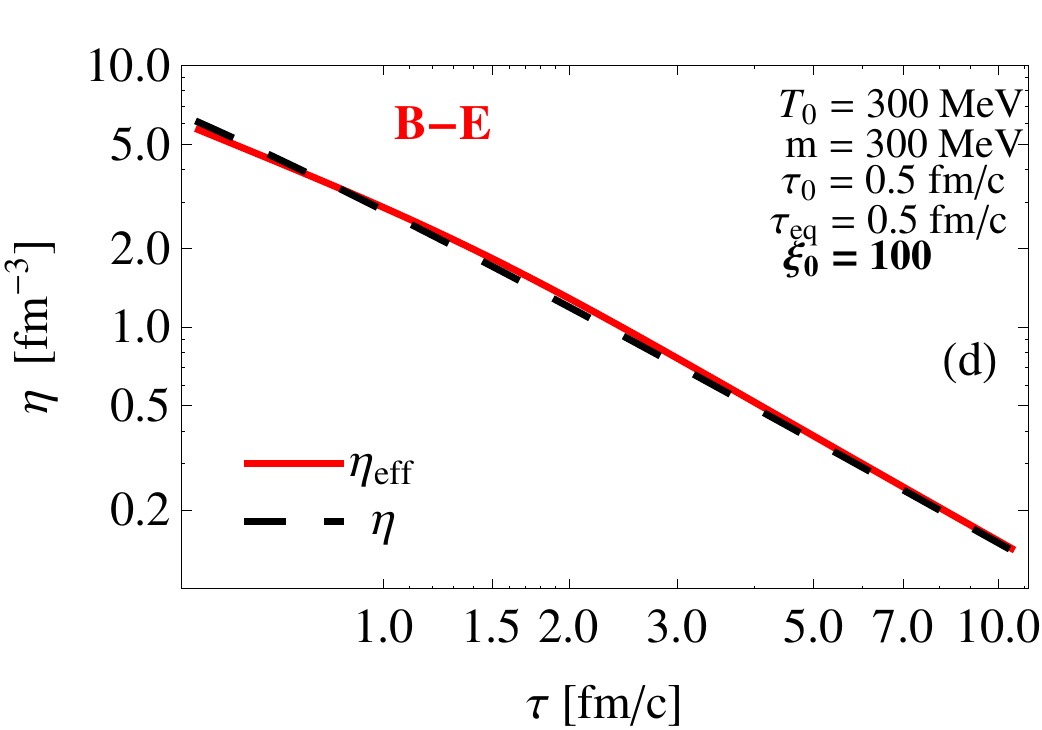}
\end{minipage}
\begin{minipage}[b]{5.05cm}
\centering
\includegraphics[angle=0,width=1\textwidth]{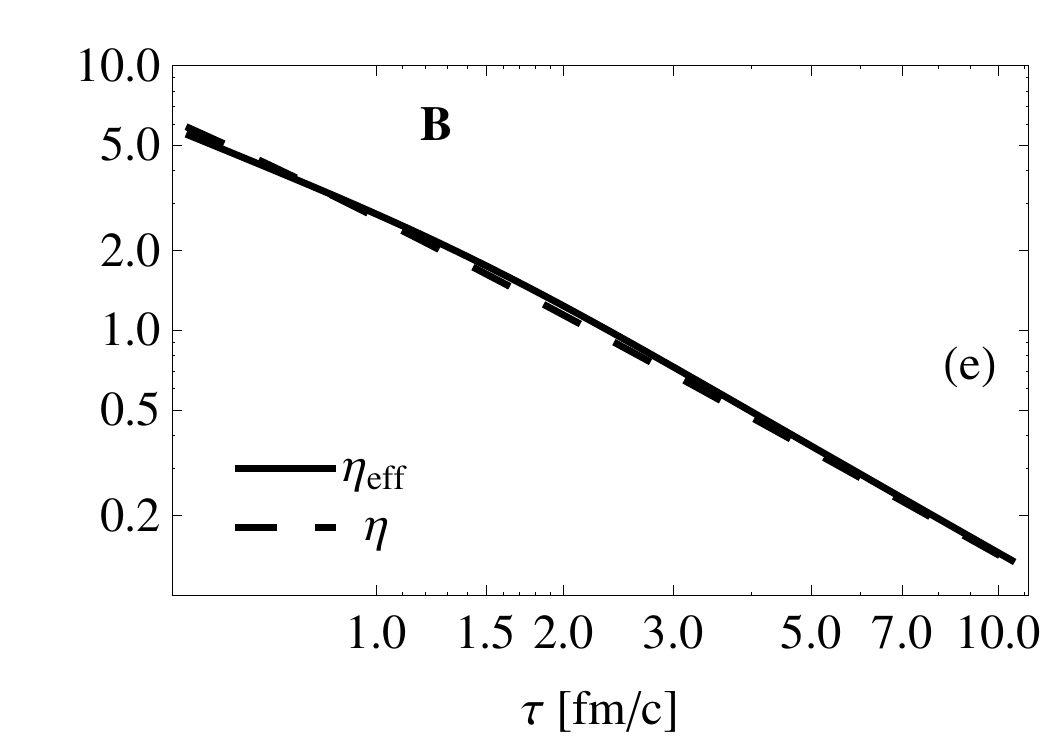}
\end{minipage}
\begin{minipage}[b]{5.05cm}
\centering
\includegraphics[angle=0,width=1\textwidth]{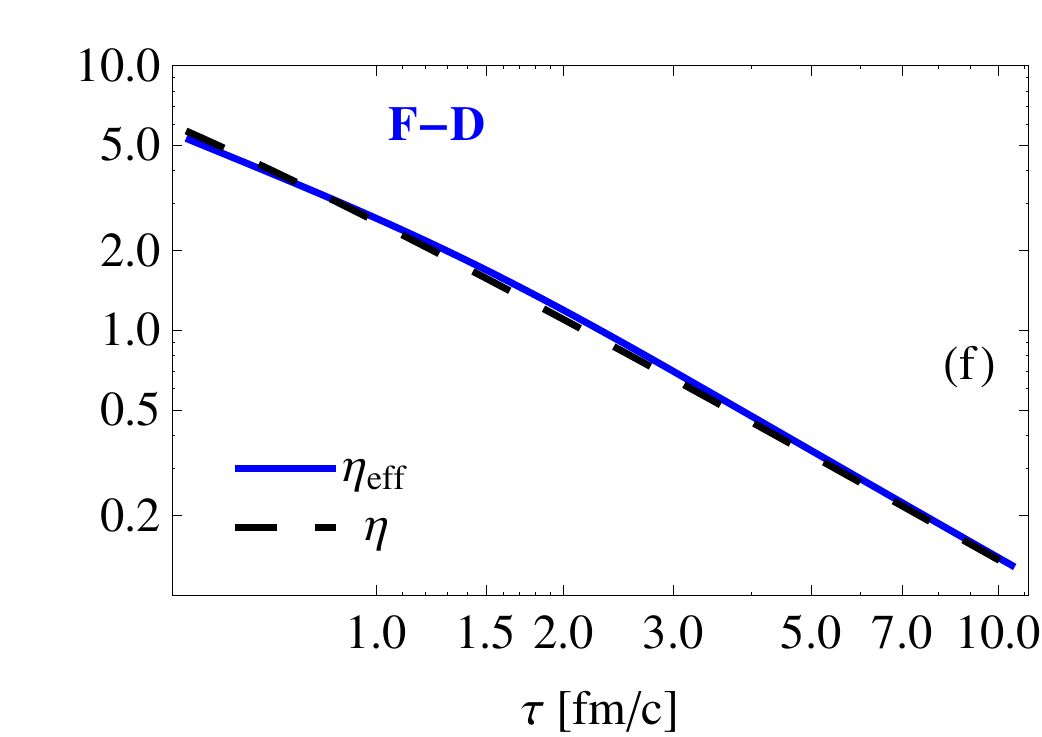}
\end{minipage}
\caption{Time dependence of the shear viscosity calculated from Eqs.~(\ref{eta_qs}) and (\ref{etakin}).}
\label{fig:Shear}
\end{center}
\end{figure}

In Fig.~\ref{fig:Shear} we show the time dependence of the shear viscosity calculated in two ways, according to Eqs.~(\ref{eta_qs}) and (\ref{etakin}). The results obtained from the exact solution of the kinetic theory are shown as the solid lines, while the results based on the expression (\ref{eta_qs}) are represented by the dashed lines. The upper row, panels (a)--(c), shows the results for the initial condition corresponding to the isotropic state, while the lower row, panels (d)--(f), shows the results for the initial state which is highly oblate. The three columns show our results for three different statistics; Bose-Einstein (B--E), Boltzmann (B), and Fermi-Dirac (F--D), from left to right, respectively. The convergence of the solid and dashed lines at late times, in all studied cases,  means that the form of the kinetic coefficient (\ref{eta_qs}) is supported by the exact solutions of the kinetic equation we have constructed in this work.

%%%%%%%%%%%%%%%%%%%%%%%%%%%%%%%%%%%%%%%%%%%%%%%%%%%%%%%%
%%%%%%%%%%%%%%%%%%%%%%%%%%%%%%%%%%%%%%%%%%%%%%%%%%%%%%%%
\begin{figure}[t]
\begin{center}
\begin{minipage}[b]{5.05cm}
\centering
\includegraphics[angle=0,width=1.01\textwidth]{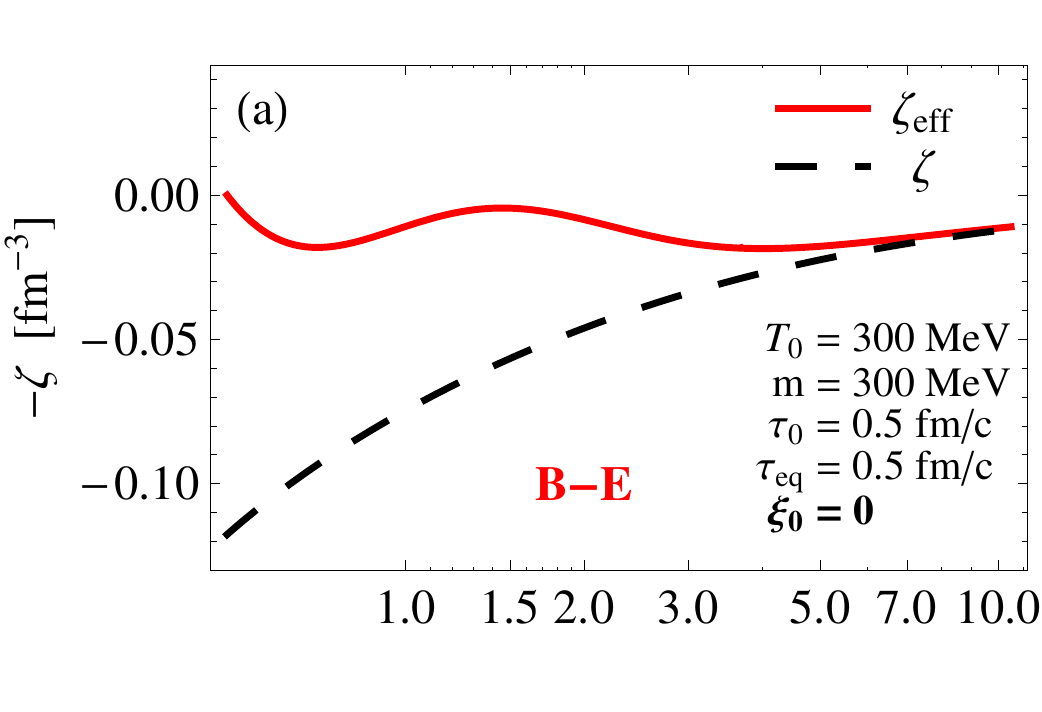}
\end{minipage}
\begin{minipage}[b]{5.05cm}
\centering
\includegraphics[angle=0,width=1\textwidth]{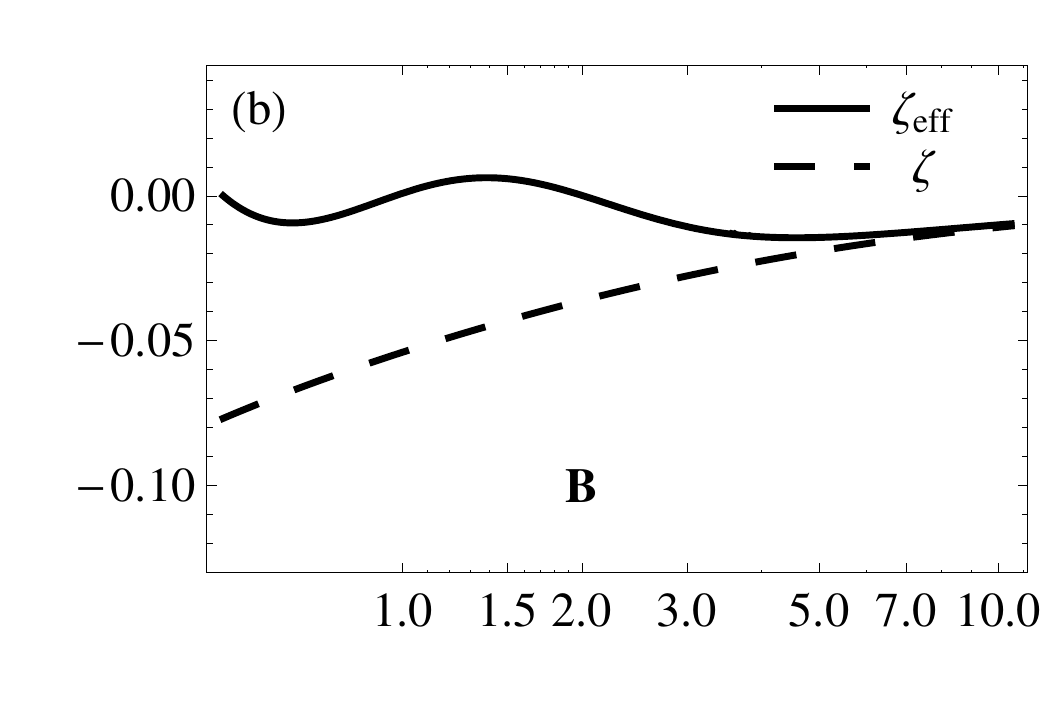}
\end{minipage}
\vspace{-0.35cm}
\begin{minipage}[b]{5.05cm}
\centering
\includegraphics[angle=0,width=1\textwidth]{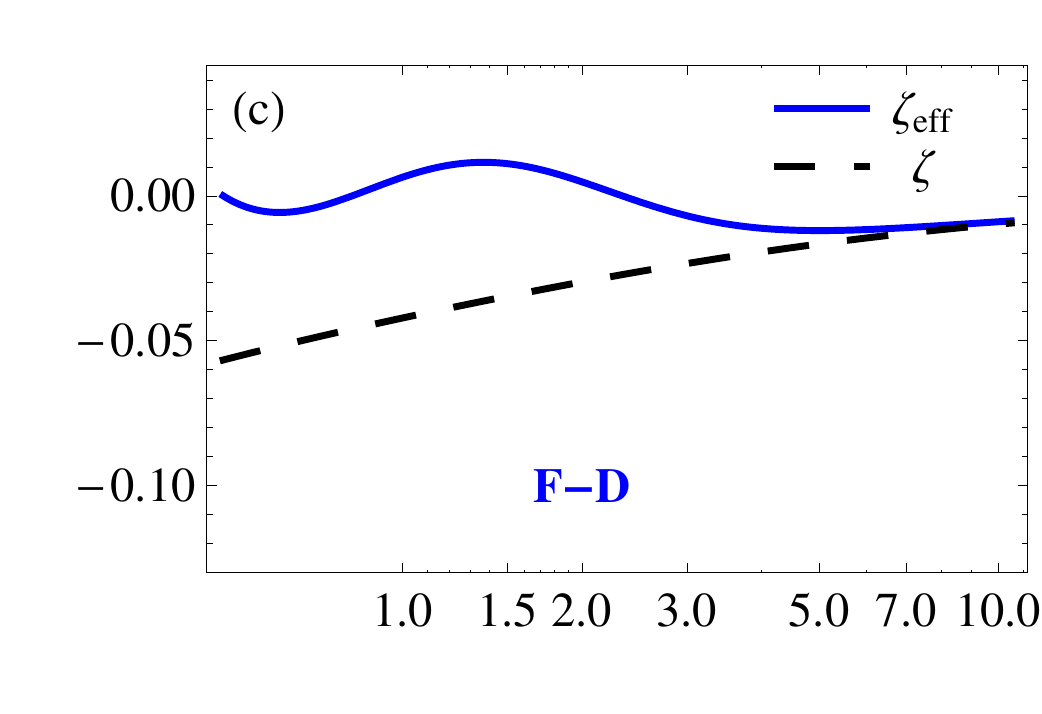}
\end{minipage}
\begin{minipage}[b]{5.05cm}
\centering
\includegraphics[angle=0,width=1\textwidth]{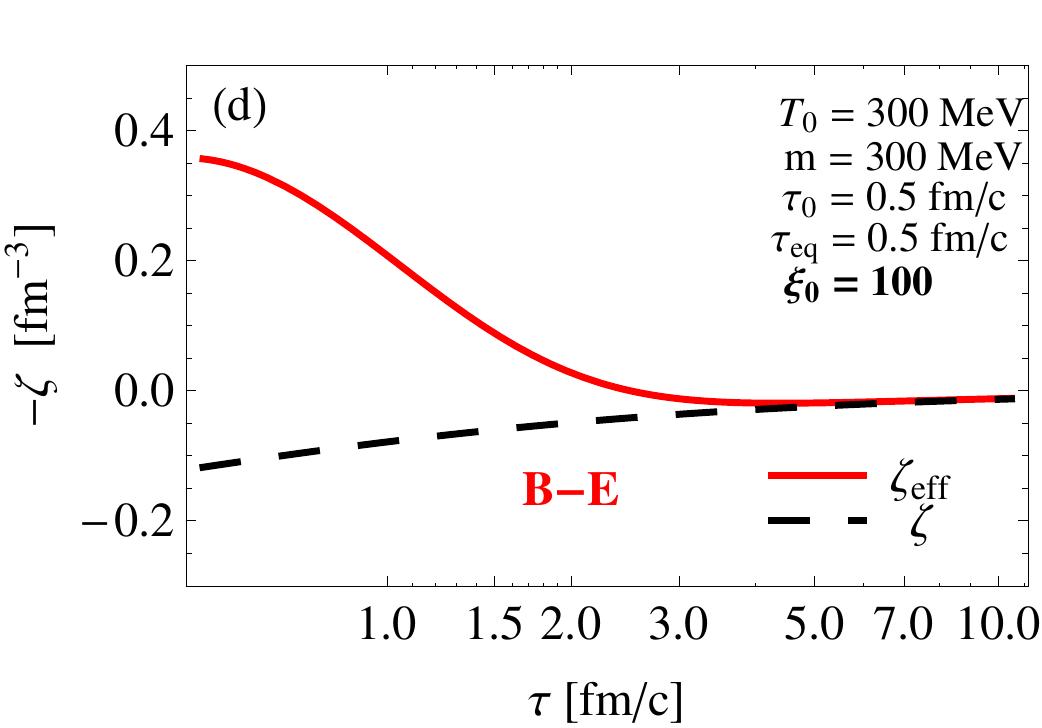}
\end{minipage}
\begin{minipage}[b]{5.05cm}
\centering
\includegraphics[angle=0,width=1\textwidth]{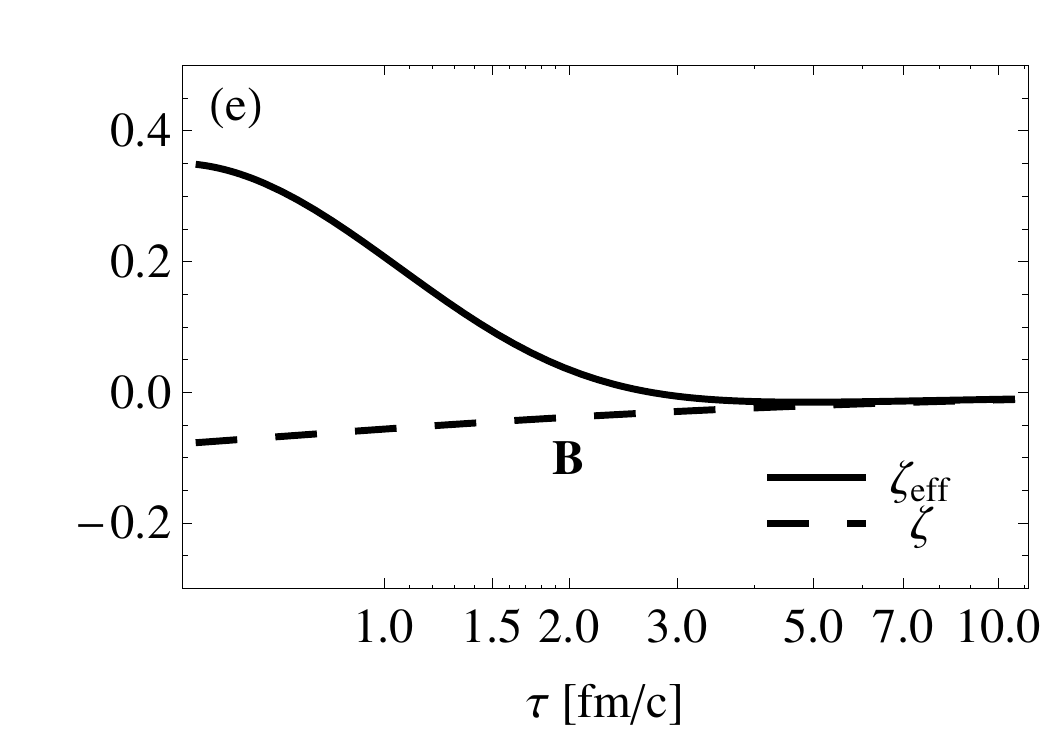}
\end{minipage}
\begin{minipage}[b]{5.05cm}
\centering
\includegraphics[angle=0,width=1\textwidth]{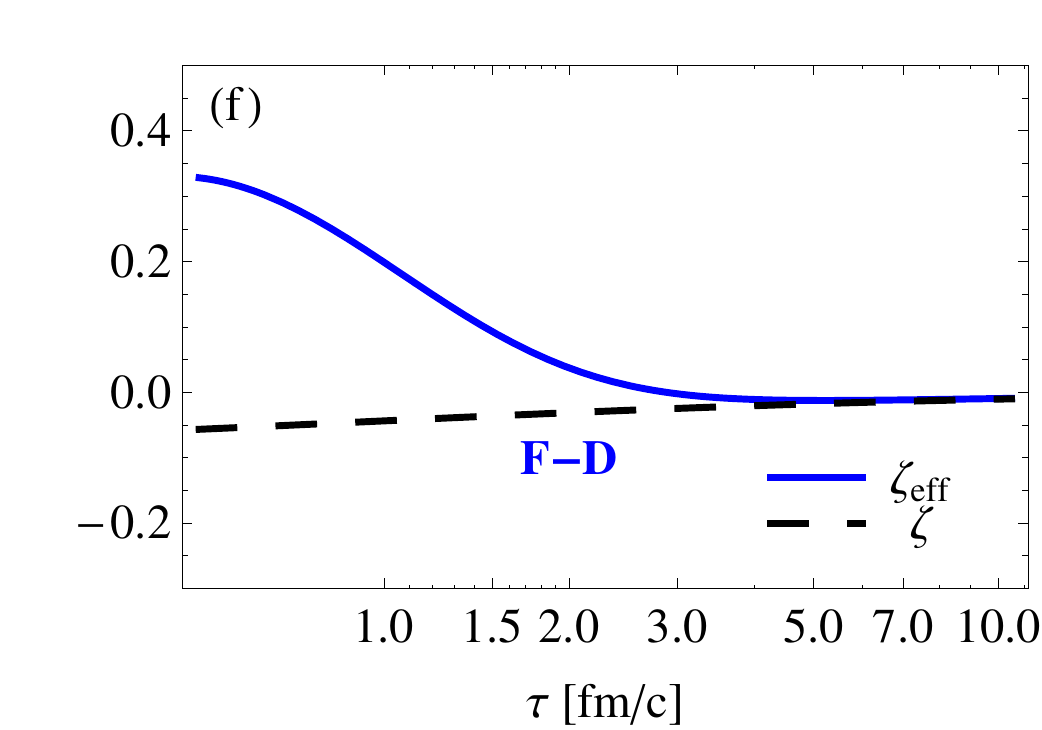}
\end{minipage}
\caption{Same as Fig.~\ref{fig:Shear} but for the bulk viscosity.}
\label{fig:Bulk}
\end{center}
\end{figure}

In Fig.~\ref{fig:Bulk} we show the time dependence of the bulk viscosity. Our results are presented in the same way as in Fig.~\ref{fig:Shear} --- the exact results based on Eq.~(\ref{zetakin}) are represented by the solid lines, while the results obtained with Eq.~(\ref{zeta_qs}) are shown as the dashed lines. Here again we find the agreement between the results based on Eqs.~(\ref{eta_qs}) and (\ref{etakin}) at late times, which confirms the validity of the formula for the kinetic coefficient (\ref{zeta_qs}).

\begin{figure}[t]
\begin{center}
\begin{minipage}[b]{6.05cm}
\centering
\includegraphics[angle=0,width=0.94\textwidth]{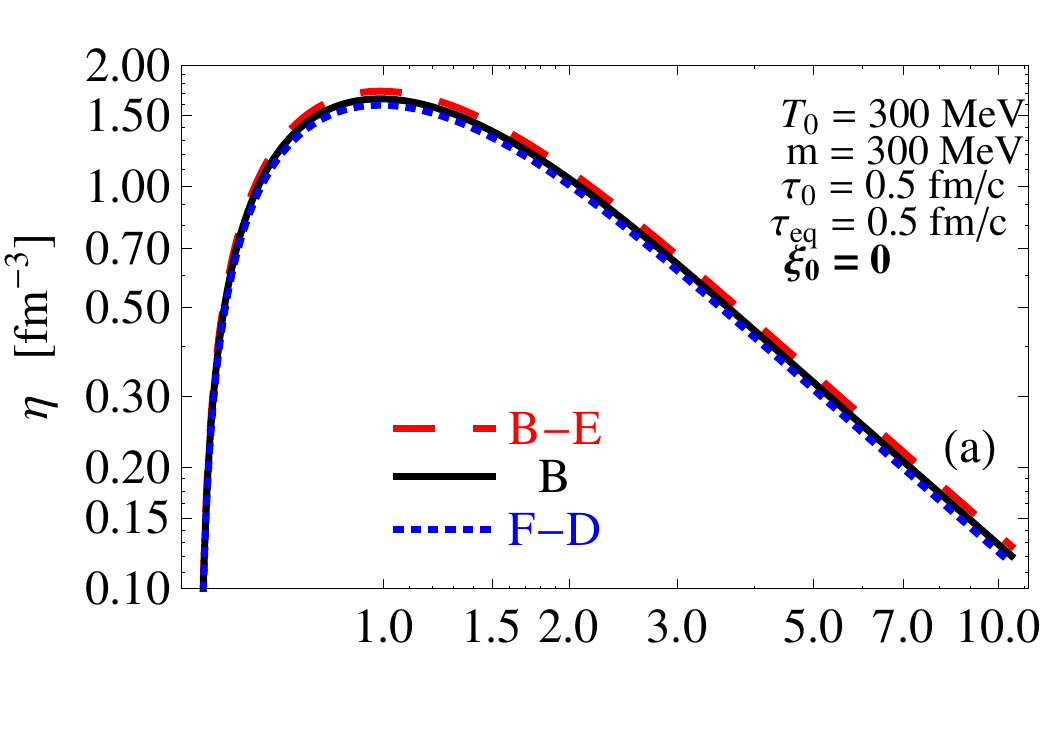}
\end{minipage}
\vspace{-0.1cm}
\begin{minipage}[b]{6.05cm}
\centering
\includegraphics[angle=0,width=0.9\textwidth]{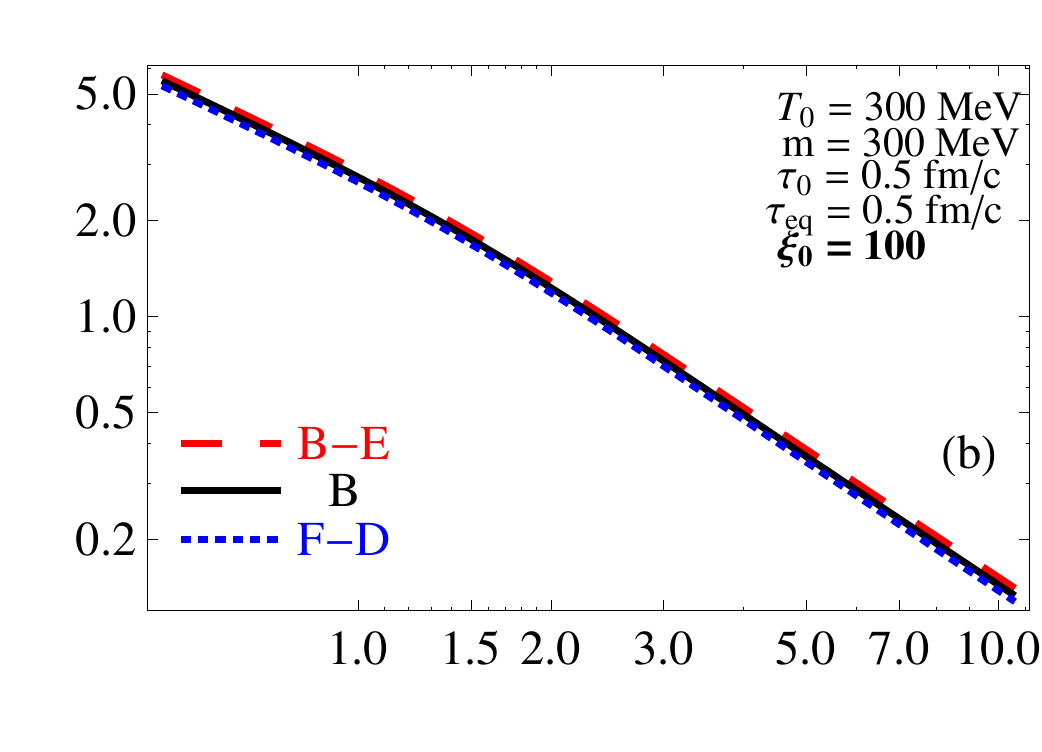}
\end{minipage}
\\
\begin{minipage}[b]{6.05cm}
\centering
\includegraphics[angle=0,width=0.94\textwidth]{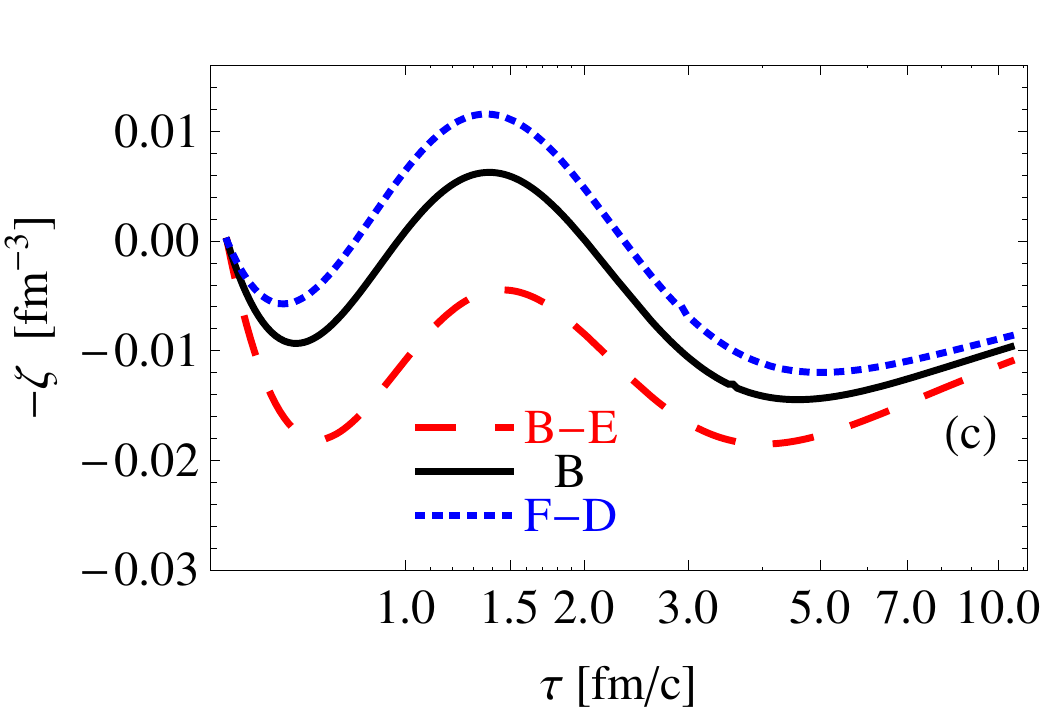}
\end{minipage}
\begin{minipage}[b]{6.05cm}
\centering
\includegraphics[angle=0,width=0.9\textwidth]{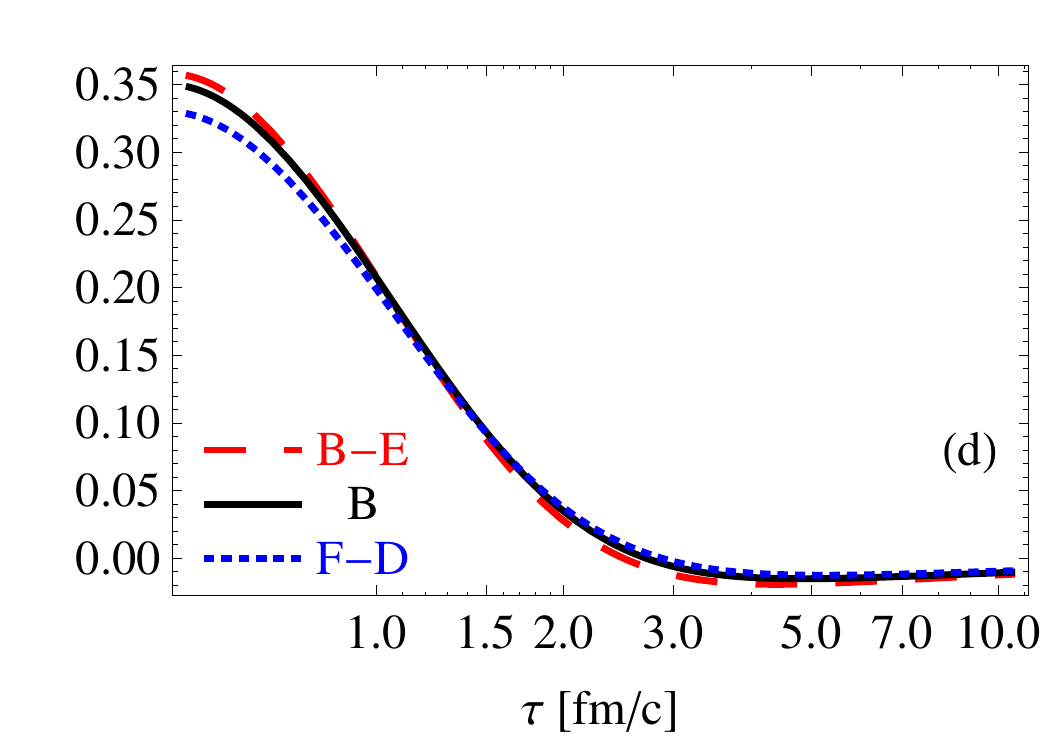}
\end{minipage}
\caption{Comparison of exact solutions for different quantum statistics.}
\label{fig:Quantum_statistics_Comparison}
\end{center}
\end{figure}

In order to compare the results obtained for different statistics in more detail, in Fig.~\ref{fig:Quantum_statistics_Comparison} we show the time dependence of the shear and bulk viscosity in the same plots. In the two upper panels, (a) and (b), we show the results for the shear viscosity, while in the two lower panels, (c) and (d), we show the bulk viscosity. In all the cases we use the exact results based on (\ref{etakin}) and (\ref{zetakin}). In the case of the shear viscosity we find only small differences between the results obtained for different statistics. On the other hand, in the case of the bulk viscosity, we find noticeable differences, especially, for the systems which are initially isotropic. This suggests that quantum forms of the kinetic coefficients should be used in the hydrodynamics calculations of higher orders than one, which may try to reproduce the time behaviour of the bulk viscous pressure also at earlier times.

%%%%%%%%%%%%%%%%%%%%%%%%%%%%%%%%%%%%%%%%%%%%%%%%%%%%%%%%%%%
%%%%%%%%%%%%%%%%%%%%%%%%%%%%%%%%%%%%%%%%%%%%%%%%%%%%%%%%%%%

%%%%%%%%%%%%%%%%%%%%%%%%%%%%%%%%%%%%%%%%%%%%%%%%%%%%%%%%%%%%%%%%%%%%%%%%%%%%%%%%%%%%%%%%%%%%%%%%%%%%
\section{Conclusions}
\label{sect:concl}
%%%%%%%%%%%%%%%%%%%%%%%%%%%%%%%%%%%%%%%%%%%%%%%%%%%%%%%%%%%%%%%%%%%%%%%%%%%%%%%%%%%%%%%%%%%%%%%%%%%%%

We have presented the exact solution of the (0+1)-dimensional Boltzmann equation for massive quantum gases. For typical initial conditions used in relativistic heavy-ion collisions, we find that the effects of quantum statistics are very small for thermodynamics-like quantities and shear viscosity. On the other hand, the quantum statistics becomes important for description of the effects connected with the bulk viscosity.

\ack

We thank Radoslaw Ryblewski and Michael Strickland for clarifying discussions and interesting comments.  This work has been supported by Polish National Science Center grant No. DEC-2012/06/A/ST2/00390. 

\appendix
%%%%%%%%%%%%%%%%%%%%%%%%%%%%%%%%%%%%%%%%%%%%%%%%%%%%%%%%%%%%%%%%%%%%%%%%%%%%%%%%%%%%%%%%%%%%%%%%%%%%
\section*{References}
\label{sect:app1}
\setcounter{section}{1}
%%%%%%%%%%%%%%%%%%%%%%%%%%%%%%%%%%%%%%%%%%%%%%%%%%%%%%%%%%%%%%%%%%%%%%%%%%%%%%%%%%%%%%%%%%%%%%%%%%%%%

%%%%%%%%%%%%%%%%%%%%%%%%%%%%%%%%%%%%%%%%%%%%%%%%%%%%%%%%%%%%%%%%%%%%%%%%%%%%%%%%%%%%%%%%%%%%%%%%%%%
%\bibliographystyle{plain}
\bibliography{qlm}

\begin{thebibliography}{10}
\expandafter\ifx\csname url\endcsname\relax
  \def\url#1{\texttt{#1}}\fi
\expandafter\ifx\csname urlprefix\endcsname\relax\def\urlprefix{URL }\fi

\bibitem{Florkowski:2014sfa}
W.~Florkowski, E.~Maksymiuk, R.~Ryblewski, M.~Strickland, Physical Review, {\bf
  C89} (2014) 054908, 1402.7348.

\bibitem{Florkowski:2013lza}
W.~Florkowski, R.~Ryblewski, M.~Strickland, Nuclear Physics, {\bf A916} (2013)
  249--259, 1304.0665.

\bibitem{Florkowski:2013lya}
W.~Florkowski, R.~Ryblewski, M.~Strickland, Physical Review, {\bf C88} (2013)
  024903, 1305.7234.

\bibitem{Florkowski:2014bba}
W.~Florkowski, R.~Ryblewski, M.~Strickland, L.~Tinti, Physical Review, {\bf
  C89} (2014) 054909, 1403.1223.

\bibitem{Nopoush:2014pfa}
M.~Nopoush, R.~Ryblewski, M.~Strickland, Physical Review, {\bf C90} (2014)
  014908, 1405.1355.

\bibitem{Denicol:2014xca}
G.~S. Denicol, U.~W. Heinz, M.~Martinez, J.~Noronha, M.~Strickland, 1408.5646.

\bibitem{Denicol:2014tha}
G.~S. Denicol, U.~W. Heinz, M.~Martinez, J.~Noronha, M.~Strickland, 1408.7048.

\bibitem{Nopoush:2014qba}
M.~Nopoush, R.~Ryblewski, M.~Strickland, 1410.6790.

\bibitem{Denicol:2014mca}
G.~S. Denicol, W.~Florkowski, R.~Ryblewski, M.~Strickland, Physical Review,
  {\bf C90} (2014) 044905, 1407.4767.

\bibitem{Jaiswal:2014isa}
A.~Jaiswal, R.~Ryblewski, M.~Strickland, Physical Review, {\bf C90} (2014)
  044908, 1407.7231.

\bibitem{Denicol:2014vaa}
G.~Denicol, S.~Jeon, C.~Gale, 1403.0962.

\bibitem{Noronha-Hostler:2013gga}
J.~Noronha-Hostler, G.~S. Denicol, J.~Noronha, R.~P.~G. Andrade, F.~Grassi,
  Physical Review, {\bf C88} (2013) 044916, 1305.1981.

\bibitem{Noronha-Hostler:2013hsa}
J.~Noronha-Hostler, G.~S. Denicol, J.~Noronha, R.~P. Andrade, F.~Grassi,
  Journal of Physics: Conference Series, {\bf 458} (2013) 012018.

\bibitem{Rose:2014fba}
J.~B. Rose, J.~F. Paquet, G.~Denicol, M.~Luzum, B.~Schenke, {\it et~al.},
  1408.0024.

\bibitem{1954PhRv...94..511B}
P.~L. {Bhatnagar}, E.~P. {Gross}, M.~{Krook}, Physical Review, {\bf 94} (1954)
  511--525.

\bibitem{Anderson:1974}
J.~Anderson, H.~Witting, Physica, {\bf 74}~(3) (1974) 466 -- 488.

\bibitem{Czyz:1986mr}
W.~Czy\ifmmode~\dot{z}\else \.{z}\fi{}, W.~Florkowski, Acta Physica Polonica,
  {\bf B17} (1986) 819--837.

\bibitem{Dyrek:1986vv}
A.~Dyrek, W.~Florkowski, Physical Review, {\bf D36} (1987) 2172.

\bibitem{cerc}
C.~Cercignani, G.~Kremer, {\em The Relativistic Boltzmann Equation: Theory and
  Applications}, Progress in Mathematical Physics, Springer Verlag NY, 2002.

\bibitem{Sasaki:2008fg}
C.~Sasaki, K.~Redlich, Physical Review, {\bf C79} (2009) 055207, 0806.4745.

\bibitem{Bozek:2009dw}
P.~Bozek, Physical Review, {\bf C81} (2010) 034909, 0911.2397.

\bibitem{Romatschke:2011qp}
P.~Romatschke, Physical Review, {\bf D85} (2012) 065012, 1108.5561.

\bibitem{florkowski2010}
W.~Florkowski, {\em Phenomenology of Ultra-relativistic Heavy-ion Collisions},
  World Scientific, 2010.

\bibitem{Bjorken:1982qr}
J.~D. Bjorken, Physical Review, {\bf D27} (1983) 140--151.

\bibitem{Bialas:1984wv}
A.~Bia\l{}as, W.~Czy\ifmmode~\dot{z}\else \.{z}\fi{}, Physical Review, {\bf
  D30} (1984) 2371--2378.

\bibitem{Bialas:1987en}
A.~Bia\l{}as, W.~Czy\ifmmode~\dot{z}\else \.{z}\fi{}, Nuclear Physics, {\bf
  B296}~(3) (1988) 611 -- 624.

\bibitem{Florkowski:2011jg}
W.~Florkowski, R.~Ryblewski, Physical Review, {\bf C85} (2012) 044902,
  1111.5997.

\bibitem{Martinez:2012tu}
M.~Martinez, R.~Ryblewski, M.~Strickland, Physical Review, {\bf C85} (2012)
  064913, 1204.1473.

\bibitem{Baym:1984np}
G.~Baym, Physics Letters, {\bf B138} (1984) 18--22.

\bibitem{Baym:1985tna}
G.~Baym, Nuclear Physics, {\bf A418} (1984) 525C--537C.

\bibitem{Heiselberg:1995sh}
H.~Heiselberg, X.-N. Wang, Physical Review, {\bf C53} (1996) 1892--1902,
  hep-ph/9504244.

\bibitem{Wong:1996va}
S.~Wong, Physical Review, {\bf C54} (1996) 2588--2599, hep-ph/9609287.

\bibitem{Romatschke:2003ms}
P.~Romatschke, M.~Strickland, Physical Review, {\bf D68} (2003) 036004,
  hep-ph/0304092.

\bibitem{Banerjee:1989by}
B.~Banerjee, R.~Bhalerao, V.~Ravishankar, Physics Letters, {\bf B224} (1989)
  16.

\end{thebibliography}
%%%%%%%%%%%%%%%%%%%%%%%%%%%%%%%%%%%%%%%%%%%%%%%%%%%%%%%%%%%%%%%%%%%%%%%%%%%%%%%%%%%%%%%%%%%%%%%%%%%

\end{document}